\documentclass[10pt,a4paper,english]{IEEEtran}
\usepackage[T1]{fontenc}
\usepackage[latin9]{inputenc}
\usepackage{color}
\usepackage{babel}
\usepackage{verbatim}
\usepackage{float}
\usepackage{mathrsfs}
\usepackage{amsthm}
\usepackage{amsmath}
\usepackage{amssymb}
\usepackage{graphicx}
\usepackage{setspace}
\usepackage{esint}
\usepackage[unicode=true,pdfusetitle,
 bookmarks=false,
 breaklinks=false,pdfborder={0 0 1},backref=false,colorlinks=true]
 {hyperref}
\hypersetup{
 linkcolor=blue, citecolor=red}

\makeatletter

\pdfpageheight\paperheight
\pdfpagewidth\paperwidth

\floatstyle{ruled}
\newfloat{algorithm}{tbp}{loa}
\providecommand{\algorithmname}{Algorithm}
\floatname{algorithm}{\protect\algorithmname}

\theoremstyle{plain}
\newtheorem{thm}{\protect\theoremname}
\theoremstyle{plain}
\newtheorem{lem}[thm]{\protect\lemmaname}

\usepackage{subfigure}
\usepackage{epstopdf}

\makeatother

\providecommand{\lemmaname}{Lemma}
\providecommand{\theoremname}{Theorem}

\begin{document}
% paper title

\title{Performance Impact of LoS and NLoS Transmissions in Dense Cellular
Networks% <-this % stops a space
}

\author{\noindent {\normalsize{}Ming Ding, }\textit{\normalsize{}Member,
IEEE}{\normalsize{}, Peng Wang, }\textit{\normalsize{}Student Member,
IEEE}{\normalsize{}, David L$\acute{\textrm{o}}$pez-P$\acute{\textrm{e}}$rez,
}\textit{\normalsize{}Member, IEEE}{\normalsize{}, \protect \\ Guoqiang Mao, }\textit{\normalsize{}Senior
Member, IEEE}{\normalsize{}, Zihuai Lin, }\textit{\normalsize{}Senior
Member, IEEE}%
%\thanks{Paper approved by xxx xxx. Manuscript received on xxx xx, 2015;
%revised on xxx xx, 2015, xxx xx, 2015, and xxx xx, 2015.}
\thanks{Ming Ding is with Data 61, Australia (e-mail: Ming.Ding@nicta.com.au).}
%\protect \\
\thanks{Peng Wang is with the School of Electrical and Information Engineering,
The University of Sydney, Australia and Data 61, Australia (e-mail:
thomaspeng.wang@sydney.edu.au).}
%\protect \\
\thanks{David L$\acute{\textrm{o}}$pez-P$\acute{\textrm{e}}$rez is with
Bell Labs Alcatel-Lucent, Ireland (email: dr.david.lopez@ieee.org).}
%\protect \\
\thanks{Guoqiang Mao is with the School of Computing and Communication, The
University of Technology Sydney, Australia and Data 61, Australia
(e-mail: g.mao@ieee.org).}
%\protect \\
\thanks{Zihuai Lin is with the School of Electrical and Information Engineering,
The University of Sydney, Australia (e-mail: zihuai.lin@sydney.edu.au).%
}% <-this % stops a space
%\thanks{Preliminary results related to this paper were presented at IEEE Globecom Conference 2015~\cite{our_GC_paper_2015_HPPP}.}% <-this % stops a space
\thanks{1536-1276 © 2015 IEEE. Personal use is permitted, but republication/redistribution requires IEEE permission. Please find the final version in IEEE from the link: http://ieeexplore.ieee.org/xpl/articleDetails.jsp?arnumber=7335646. Digital Object Identifier 10.1109/TWC.2015.2503391}
}

\maketitle
\begin{abstract}
In this paper, we introduce a sophisticated path loss model incorporating both line-of-sight (LoS) and non-line-of-sight (NLoS) transmissions to study their impact on the performance of dense small cell networks (SCNs).
Analytical results are obtained for the coverage probability and the area spectral efficiency  (ASE),
assuming both a general path loss model and a special case with a linear LoS probability function.
The performance impact of LoS and NLoS transmissions in dense SCNs in terms of the coverage probability and  the ASE is significant,
both quantitatively and qualitatively,
compared with the previous work that does not differentiate LoS and NLoS transmissions.
Our analysis demonstrates that the network coverage probability first increases with the increase of the base station (BS) density,
and then decreases as the SCN becomes denser.
This decrease further makes the ASE suffer from a slow growth or even a decrease with network densification.
The ASE will grow almost linearly as the BS density goes ultra dense.
For practical regime of the BS density,
the performance results derived from our analysis are distinctively different from previous results,
and thus shed new insights on the design and deployment of future dense SCNs.
\end{abstract}

\begin{IEEEkeywords}
stochastic geometry, Homogeneous Poisson Point Process (HPPP), Line-of-Sight
(LoS), Non-Line-of-Sight (NLoS), dense small cell networks (SCNs),
coverage probability, area spectral efficiency (ASE).
\end{IEEEkeywords}

\section{Introduction\label{sec:Introduction}}

Driven by a new generation of wireless user equipment (UE) and the proliferation of bandwidth-intensive applications,
mobile data traffic and network load are increasing in an exponential manner,
and are straining current cellular networks to a breaking point~\cite{Report_CISCO}.
In this context, small cell networks (SCNs),
comprised of remote radio heads, metrocells, picocells, and/or femtocells,
have attracted much attention as one of the most promising approaches to increase network capacity and meet the ever-increasing capacity demands~\cite{Tutor_smallcell,TR36.872}.
SCNs can significantly enhance network capacity through a high spatial spectrum reuse,
e.g., network capacity could potentially grow linearly with the number of small cells~\cite{Tutor_smallcell}.
Due to such capacity gains,
the orthogonal deployment of SCNs within the existing macrocell network,
i.e., small cells and macrocells operate on different frequency spectrum (Small Cell Scenario \#2a defined in~\cite{TR36.872}),
have gained much momentum in the design of the 4th-generation (4G) Long Term Evolution (LTE) networks by the 3rd Generation Partnership Project (3GPP).
Orthogonal deployments of dense SCNs are also envisaged as the workhorse for capacity enhancement in the 5th-generation (5G) networks,
aided by its easy deployment arising from its low interaction with the macrocell tier,
e.g., no inter-tier interference~\cite{Tutor_smallcell}.
In this paper, our focus is on these orthogonal deployments of dense SCNs.

In order to deploy dense SCNs in a cost-effective manner,
vendors and operators need foremost a deep theoretical understanding of the implications that small cells bring about.
Being aware of the need for such knowledge,
the wireless industry and research community have been working relentlessly on the modeling and the analysis of the dense SCN deployments.
However, up to now, most studies on SCNs have considered only simplistic path loss models that do not differentiate Line-of-Sight (LoS) and Non-Line-of-Sight (NLoS) transmissions {[}4-7{]}.
It is well known that LoS transmission may occur when the distance between a transmitter and a receiver is small,
and NLoS transmission is common in office environments and in central business districts.
Moreover, when the distance between a transmitter and a receiver decreases,
the probability that a LoS path exists between them increases,
thereby causing a transition from NLoS transmission to LoS transmission with a higher probability.
In this paper, we will study the performance impact of such NLoS-to-LoS transition in dense SCNs.
The main contributions of this paper are as follows:
\begin{itemize}
\item
Analytical results are obtained for the network performance of the coverage probability and the area spectral efficiency (ASE), using a proposed general path loss model incorporating both LoS and NLoS transmissions.
\item
Using the above results,
numerically tractable integral-form expressions for the coverage probability and the ASE are further obtained for a 3GPP path loss model with a \emph{linear} LoS probability function.
Our analysis can be readily extended to deal with more complicated path loss models by approximating the corresponding LoS probability function as a piece-wise \emph{linear} function.
\item
Our theoretical analysis reveals an important finding,
i.e., when the density of small cell base station (BSs) is larger than a threshold,
the network coverage probability will decrease as small cells become denser,
which in turn can make the ASE suffer from a slow growth or even a decrease.
Thereafter, the ASE will grow almost linearly as the small cell BS density increases above another threshold,
larger than the previous one.
This finding that the ASE performance may suffer from a slow growth or even a decrease as the BS density increases,
is not only quantitatively but also qualitatively different from previous study results with a simplistic path loss model that does not differentiate LoS and NLoS transmissions.
Although our conclusion is made from the investigated set of parameters recommended by the 3GPP for SCNs,
which sheds valuable insights on the design and deployment of practical SCNs,
it is of significant interest to further study the generality of our conclusion in other network models and with other parameter sets.
\end{itemize}

The remainder of this paper is structured as follows.
Section~\ref{sec:Related-Work} provides a brief review on stochastic geometry and compares the closest related works to our work.
Section~\ref{sec:System-Model} describes the system model.
Section~\ref{sec:General-Results} presents our main analytical results on the coverage probability and the ASE,
followed by their application in a 3GPP study case addressed in Section~\ref{sec:A-3GPP-Special-Case}.
The numerical results are discussed in Section~\ref{sec:Simulation-and-Discussion},
with remarks shedding some new light on the performance and the deployment of dense SCNs.
Finally, the conclusions are drawn in Section~\ref{sec:Conclusion}.

\section{Related Work\label{sec:Related-Work}}

In stochastic geometry,
BS positions are typically modeled as a Homogeneous Poisson Point Process (HPPP) on the plane,
and closed-form expressions of coverage probability can be found for some scenarios in single-tier cellular networks~\cite{Jeff's work 2011} and multi-tier cellular networks~\cite{JSAC_Dhillon,TWC_Singh}.
A general treatment of stochastic geometry can be found in~\cite{book_Haenggi}.
The major conclusion in~{[}4-7{]} is that neither the number of cells nor the number of cell tiers changes the coverage probability in interference-limited fully-loaded wireless networks.
However, these works consider a simplistic path loss model that does not differentiate LoS and NLoS transmissions.
In contrast, in this paper, we consider a more complete path loss model incorporating both LoS and NLoS transmissions to study their impact on the performance of dense SCNs.

Notions that are similar to LoS and NLoS transmissions have been previously explored in the building blockage study in~\cite{Analysis_random_blockage} and the indoor communication network in~\cite{Wall_pene_indoor}.
In~\cite{Analysis_random_blockage},
the authors proposed a microscopic performance analysis framework to model the random blockage effect of buildings,
and analyze its impact on cellular network performance.
Further refinement and verification of the proposed model in~\cite{Analysis_random_blockage} is needed,
especially to consider reflections which are an important contributor to coverage in urban areas.
In~\cite{Wall_pene_indoor},
the authors present an analytical study of indoor propagation through walls,
and showed that the throughput does not scale linearly with the density of small cells.
Different from~\cite{Wall_pene_indoor},
in this paper, we investigate outdoor dense SCNs.

The closest related works to the one in this paper are~\cite{related_work_Jeff},~\cite{Related_work_Health} and~\cite{related_work_Galiotto}.

In~\cite{related_work_Jeff},
the authors assumed a multi-slope piece-wise path loss function.
Specifically, assuming that the distance between a BS and a UE is denoted by $r$ in km,
then the path loss associated with distance $r$ can be formulated as
%\begin{singlespace}
\begin{equation}
\zeta\left(r\right)=\begin{cases}
\begin{array}{l}
\zeta_{1}\left(r\right),\\
\zeta_{2}\left(r\right),\\
\vdots\\
\zeta_{N}\left(r\right),
\end{array} & \hspace{-0.3cm}\begin{array}{l}
\textrm{when }0\leq r\leq d_{1}\\
\textrm{when }d_{1}<r\leq d_{2}\\
\vdots\\
\textrm{when }r>d_{N-1}
\end{array}\end{cases},\label{eq:PL_model_Jeff}
\end{equation}
%\end{singlespace}
\noindent where the path loss function $\zeta\left(r\right)$ is segmented into $N$ pieces,
with each piece and each segment break point denoted by $\zeta_{n}\left(r\right)$ and $d_{n}$, respectively.
%,n\in\left\{ 1,2,\ldots,N-1\right\}

In~\cite{Related_work_Health},
the authors treated the event of LoS or NLoS transmission as a probabilistic event for a millimeter wave communication scenario.
Specifically, the path loss associated with distance $r$ is formulated as
%\begin{singlespace}
\begin{equation}
\zeta\left(r\right)=\begin{cases}
\begin{array}{l}
\zeta^{\textrm{L}}\left(r\right),\\
\zeta^{\textrm{NL}}\left(r\right),
\end{array} & \hspace{-0.3cm}\begin{array}{l}
\textrm{with probability }\textrm{Pr}^{\textrm{L}}\left(r\right)\\
\textrm{with probability }\left(1-\textrm{Pr}^{\textrm{L}}\left(r\right)\right)
\end{array}\end{cases},\label{eq:PL_model_Robert}
\end{equation}
%\end{singlespace}
\noindent where $\zeta^{\textrm{L}}\left(r\right)$, $\zeta^{\textrm{NL}}\left(r\right)$ and $\textrm{Pr}^{\textrm{L}}\left(r\right)$ are the path loss function for the case of LoS transmission, the path loss function for the case of NLoS transmission and the LoS probability function, respectively.
To simplify the analysis,
the LoS probability function $\textrm{Pr}^{\textrm{L}}\left(r\right)$
was approximated by a moment matched equivalent step function in~\cite{Related_work_Health}.

In~\cite{related_work_Galiotto},
the authors used the same path loss model as in (\ref{eq:PL_model_Robert}) and considered the approximation of $\textrm{Pr}^{\textrm{L}}\left(r\right)$ as an exponentially decreasing function.
The results in~\cite{related_work_Galiotto} are less tractable than those in~\cite{related_work_Jeff} and~\cite{Related_work_Health}.
This is because the exponentially decreasing LoS probability function,
albeit more practical than the step function in~\cite{Related_work_Health},
is still difficult to deal with in the analysis.

In this paper, we extend the works in~{[}10-12{]} to an even more general scenario to improve the following aspects.
In~\cite{related_work_Jeff},
the multi-slope piece-wise path loss model in (\ref{eq:PL_model_Jeff}) does not fit well with the practical model defined by the 3GPP,
in which the path loss function is not a one-to-one mapping to the distance.
In~\cite{Related_work_Health},
the single-piece path loss model and the proposed step function are not compatible with the practical piece-wise path loss functions,
the detailed modeling of which is presented in Section~\ref{sec:System-Model}.
In~\cite{related_work_Galiotto},
the considered path loss model is very practical.
However, the generality of the study and the tractability of the analysis need to be improved.
Compared with~{[}10-12{]}, the novelties of our paper are summarized as follows,
\begin{itemize}
\item
\textbf{Extending the existing works to a more general scenario: }
We propose a path loss model that features piece-wise path loss functions with probabilistic LoS and NLoS transmissions.
The proposed path loss model, to be presented in Section~\ref{sec:System-Model}, is general and can be applied to several channel models that capture LoS and NLoS transmissions~{[}10-14{]}.
%includes the path loss models in~{[}10-12{]} as its special cases.
Moreover, new analytical results are obtained for the coverage probability and the ASE.
%The proposed path loss model to be presented in Section~\ref{sec:System-Model} is general and can be applied to several channel models that capture LoS and NLoS transmissions~{[}10-12{]}.
\item
\noindent
\textbf{Presenting more tractable results for a special path loss model with a }\textbf{\emph{linear}}\textbf{ LoS probability function:}
We derive numerically tractable integral-form expressions for the coverage probability and the ASE for a 3GPP path loss model with a \emph{linear} LoS probability function.
The results are more simple and tractable than those in~\cite{related_work_Galiotto},
and no approximation such as the step function~\cite{Related_work_Health} is used in our analysis.
Although the results in~\cite{related_work_Jeff} are superior than ours in terms of tractability,
our analysis can characterize practical networks more accurately due to the consideration of probabilistic LoS events and a linear LoS probability function.
Note that the inclusion of a linear LoS probability function is not trivial compared with~\cite{related_work_Jeff} in the sense that,
in our model a UE may be associated with a BS that is not the nearest BS to the UE,
but where such BS has a LoS path to the UE resulting in the smallest path loss (i.e., with the largest $\zeta\left(r\right)$).
In contrast, in~\cite{related_work_Jeff},
a UE always connects with its nearest BS.
%a UE is assumed to be always connected with its nearest BS.
Moreover, our study on the path loss model with the \emph{linear} LoS probability function can be extended to various cases.
In this paper, we show that the approach of approximating a LoS probability function by a piece-wise \emph{linear} function and invoking our results,
can deal with complicated path loss models in a tractable manner.
\item
\textbf{Disclosing a new finding on the ASE performance: }
We present a new finding that the ASE performance without the assumption of near-field path loss exponents may suffer from a slow growth or even a decrease with BS densification.
This new finding is not only quantitatively but also qualitatively different from the results given by~{[}10-12{]}.
Note that in~\cite{related_work_Jeff},
a deterministic rate based on the signal to interference plus noise ratio (SINR) threshold is assumed for the typical UE,
no matter what the actual SINR value is.
Our definition of the ASE, to be formally presented in Section~\ref{sec:System-Model}, considers a more realistic SINR-dependent rate,
which leads to a more complex  analysis that requires one more fold of numerical integral compared with~\cite{related_work_Jeff}.
\end{itemize}

\section{System Model\label{sec:System-Model}}

We consider a downlink (DL) cellular network with BSs deployed in a plane according to an HPPP $\Phi$ of intensity $\lambda$ $\textrm{BSs/km}^{2}$.
UEs are Poisson distributed in the considered network with an intensity of $\lambda^{\textrm{UE}}$ $\textrm{UEs/km}^{2}$.
Note that $\lambda^{\textrm{UE}}$ is assumed to be sufficiently larger than $\lambda$
so that each BS has at least one associated UE in its coverage.
As in (\ref{eq:PL_model_Jeff}) and (\ref{eq:PL_model_Robert}),
the distance between an arbitrary BS and an arbitrary UE is denoted by $r$ in km.
Considering practical LoS and NLoS transmissions,
we propose to model the path loss with respect to distance $r$ as (\ref{eq:prop_PL_model}),
which is shown on the top of next page.
\begin{algorithm*}
%\begin{singlespace}
%\noindent
\begin{equation}
\zeta\left(r\right)=\begin{cases}
\zeta_{1}\left(r\right)=\begin{cases}
\begin{array}{l}
\zeta_{1}^{\textrm{L}}\left(r\right),\\
\zeta_{1}^{\textrm{NL}}\left(r\right),
\end{array} & \hspace{-0.3cm}\begin{array}{l}
\textrm{with probability }\textrm{Pr}_{1}^{\textrm{L}}\left(r\right)\\
\textrm{with probability }\left(1-\textrm{Pr}_{1}^{\textrm{L}}\left(r\right)\right)
\end{array}\end{cases}\hspace{-0.3cm}, & \textrm{when }0\leq r\leq d_{1}\\
\zeta_{2}\left(r\right)=\begin{cases}
\begin{array}{l}
\zeta_{2}^{\textrm{L}}\left(r\right),\\
\zeta_{2}^{\textrm{NL}}\left(r\right),
\end{array} & \hspace{-0.3cm}\begin{array}{l}
\textrm{with probability }\textrm{Pr}_{2}^{\textrm{L}}\left(r\right)\\
\textrm{with probability }\left(1-\textrm{Pr}_{2}^{\textrm{L}}\left(r\right)\right)
\end{array}\end{cases}\hspace{-0.3cm}, & \textrm{when }d_{1}<r\leq d_{2}\\
\vdots & \vdots\\
\zeta_{N}\left(r\right)=\begin{cases}
\begin{array}{l}
\zeta_{N}^{\textrm{L}}\left(r\right),\\
\zeta_{N}^{\textrm{NL}}\left(r\right),
\end{array} & \hspace{-0.3cm}\begin{array}{l}
\textrm{with probability }\textrm{Pr}_{N}^{\textrm{L}}\left(r\right)\\
\textrm{with probability }\left(1-\textrm{Pr}_{N}^{\textrm{L}}\left(r\right)\right)
\end{array}\end{cases}\hspace{-0.3cm}, & \textrm{when }r>d_{N-1}
\end{cases}.\label{eq:prop_PL_model}
\end{equation}
%\end{singlespace}
\end{algorithm*}

In (\ref{eq:prop_PL_model}), the path loss function $\zeta\left(r\right)$ is segmented into $N$ pieces with each piece denoted by $\zeta_{n}\left(r\right)$.
Besides, $\zeta_{n}^{\textrm{L}}\left(r\right)$, $\zeta_{n}^{\textrm{NL}}\left(r\right)$ and $\textrm{Pr}_{n}^{\textrm{L}}\left(r\right)$ are the $n$-th piece of path loss function for the LoS transmission,
the $n$-th piece of path loss function for the NLoS transmission,
and the $n$-th piece of the LoS probability function, respectively.
Note that the proposed model is general and can be applied to several channel models that capture LoS and NLoS transmissions~{[}10-14{]}.

Moreover, $\zeta_{n}^{\textrm{L}}\left(r\right)$ and $\zeta_{n}^{\textrm{NL}}\left(r\right)$ in (\ref{eq:prop_PL_model}) are modeled as
\begin{equation}
\zeta_{n}\left(r\right)=\begin{cases}
\begin{array}{l}
\zeta_{n}^{\textrm{L}}\left(r\right)=A_{n}^{{\rm {L}}}r^{-\alpha_{n}^{{\rm {L}}}},\\
\zeta_{n}^{\textrm{NL}}\left(r\right)=A_{n}^{{\rm {NL}}}r^{-\alpha_{n}^{{\rm {NL}}}},
\end{array} & \hspace{-0.3cm}\begin{array}{l}
\textrm{for LoS}\\
\textrm{for NLoS}
\end{array},\end{cases}\label{eq:PL_BS2UE}
\end{equation}
\noindent where $A_{n}^{{\rm {L}}}$ and $A_{n}^{{\rm {NL}}}, n\in\left\{ 1,2,\ldots,N\right\} $ are the path losses at a reference distance $r=1$ for the LoS and the NLoS cases in $\zeta_{n}\left(r\right)$, respectively, and $\alpha_{n}^{{\rm {L}}}$ and $\alpha_{n}^{{\rm {NL}}}, n\in\left\{ 1,2,\ldots,N\right\} $ are the path loss exponents for the LoS and the NLoS cases in $\zeta_{n}\left(r\right)$, respectively.
In practice, $A_{n}^{{\rm {L}}}$, $A_{n}^{{\rm {NL}}}$, $\alpha_{n}^{{\rm {L}}}$ and $\alpha_{n}^{{\rm {NL}}}$ are constants obtained from field tests~\cite{TR36.828},~\cite{SCM_pathloss_model}.
For convenience, $\left\{ \zeta_{n}^{\textrm{L}}\left(r\right)\right\} $ and $\left\{ \zeta_{n}^{\textrm{NL}}\left(r\right)\right\} $ are further stacked into piece-wise functions written as
%\begin{singlespace}
\begin{equation}
\zeta^{Path}\left(r\right)=\begin{cases}
\zeta_{1}^{Path}\left(r\right), & \textrm{when }0\leq r\leq d_{1}\\
\zeta_{2}^{Path}\left(r\right),\hspace{-0.3cm} & \textrm{when }d_{1}<r\leq d_{2}\\
\vdots & \vdots\\
\zeta_{N}^{Path}\left(r\right), & \textrm{when }r>d_{N-1}
\end{cases},\label{eq:general_PL_func}
\end{equation}
%\end{singlespace}
\noindent where the string variable $Path$ takes the value of ``L'' and ``NL'' for the LoS and the NLoS cases, respectively.

In (\ref{eq:prop_PL_model}),
$\textrm{Pr}_{n}^{\textrm{L}}\left(r\right),n\in\left\{ 1,2,\ldots,N\right\} $ is the $n$-th piece probability function that a transmitter and a receiver separated by a distance $r$ has a LoS path,
which is typically a monotonically decreasing function with $r$.
For convenience,
$\left\{ \textrm{Pr}_{n}^{\textrm{L}}\left(r\right)\right\} $ is further stacked into a piece-wise LoS probability function expressed as
%\begin{singlespace}
\begin{equation}
\textrm{Pr}^{\textrm{L}}\left(r\right)=\begin{cases}
\textrm{Pr}_{1}^{\textrm{L}}\left(r\right), & \textrm{when }0\leq r\leq d_{1}\\
\textrm{Pr}_{2}^{\textrm{L}}\left(r\right),\hspace{-0.3cm} & \textrm{when }d_{1}<r\leq d_{2}\\
\vdots & \vdots\\
\textrm{Pr}_{N}^{\textrm{L}}\left(r\right), & \textrm{when }r>d_{N-1}
\end{cases}.\label{eq:general_LoS_Pr}
\end{equation}
%\end{singlespace}

Our model is consistent with the ones adopted in the 3GPP~\cite{TR36.828},~\cite{SCM_pathloss_model}.
It should be noted that the considered path loss model shown in (\ref{eq:prop_PL_model}) includes the models in~{[}10-12{]} as its special cases.
More specifically, for~\cite{related_work_Jeff},
all the LoS probabilities are zero,
i.e., assuming $\textrm{Pr}_{n}^{\textrm{L}}\left(r\right)=0,\forall n\in\left\{ 1,2,\ldots,N\right\} $ in (\ref{eq:prop_PL_model}).
For~\cite{Related_work_Health} and~\cite{related_work_Galiotto},
there is only one piece of path loss function with one LoS path loss exponent and one NLoS path loss exponent,
i.e., assuming $N=1$ in (\ref{eq:prop_PL_model}).

Another important note is on the practical usage of the proposed path loss model,
which is summarized in the following:
\begin{itemize}
\item
As have been addressed in~\cite{related_work_Jeff},
the standard single-slope path loss function does not accurately capture the dependence of the path loss exponent on the link distance in many future networks,
such as the multi-ray transmission environments, the dense/clustered small cells, the millimeter wave communications with blockage, etc.
Therefore, a dual or more slope path loss function such as (\ref{eq:general_PL_func}) should be considered.
\item
The realistic LoS probability functions usually take complicated mathematical forms, e.g., in the 3GPP standards~\cite{TR36.828},
which will be addressed in more detail in Section~\ref{sec:A-3GPP-Special-Case}.
Therefore, to achieve both analytical tractability and result accuracy,
it is desirable to approximate the complicated LoS probability function as a few pieces of elementary functions, e.g., linear functions.
Such piece-wise LoS probability function is well captured by (\ref{eq:general_LoS_Pr}) and can be studied in our framework.
\item
With the justification of both (\ref{eq:general_PL_func}) and (\ref{eq:general_LoS_Pr}),
we can conclude that the proposed path loss model in (\ref{eq:prop_PL_model}) is versatile to cope with various practical network scenarios.
\end{itemize}

In this paper, we assume the following user association strategy (UAS).
Each UE should be associated with the BS having the smallest path loss (i.e., with the largest $\zeta\left(r\right)$) to the UE~\cite{Related_work_Health},~\cite{related_work_Galiotto}.
Note that in our previous work~\cite{our_GC_paper_2015_HPPP} and some existing works~\cite{Jeff's work 2011},~\cite{related_work_Jeff},
it was assumed that each UE should be connected to the BS with the closest proximity.
Such assumption is not appropriate  because in practice it is possible for a UE to associate with a BS that is not the closest one but the one with the minimum path loss.

Finally, we assume that each BS/UE is equipped with an isotropic antenna,
and as a common practice in the field~{[}4-6,10,12{]},
the multi-path fading between an arbitrary BS and an arbitrary UE is modeled as independently identical distributed (i.i.d.) Rayleigh fading.

\section{Analysis for the Proposed Path Loss Model \label{sec:General-Results}}

Using the property of the HPPP,
we study the performance of SCNs by considering the performance of a typical UE located at the origin $o$.
We first investigate the probability that the typical UE is covered by its associated BS.
This coverage probability is defined as the probability that the UE's SINR is above a per-designated threshold $\gamma$:
%\begin{singlespace}
\begin{equation}
p^{\textrm{cov}}\left(\lambda,\gamma\right)=\textrm{Pr}\left[\mathrm{SINR}>\gamma\right],\label{eq:Coverage_Prob_def}
\end{equation}
%\end{singlespace}
\noindent where the SINR is computed by
\begin{equation}
\mathrm{SINR}=\frac{P\zeta\left(r\right)h}{I_{r}+N_{0}},\label{eq:SINR}
\end{equation}
\noindent where $h$ is the channel gain of Rayleigh fading,
which is modeled as an exponential random variable (RV) with the mean of one,
and $P$ and $N_{0}$ are the transmission power of each BS and the additive white Gaussian noise (AWGN) power at each UE, respectively.
Moreover, $I_{r}$ is the aggregate interference given by
\begin{equation}
I_{r}=\sum_{i:\, b_{i}\in\Phi\setminus b_{o}}P\beta_{i}g_{i},\label{eq:cumulative_interference}
\end{equation}
\noindent where $b_{o}$ is the BS serving the typical UE,
which is  located at distance $r$ from the typical UE,
and  $b_{i}$, $\beta_{i}$ and $g_{i}$ are the $i$-th interfering BS, the path loss associated with $b_{i}$ and the multi-path fading channel gain associated with $b_{i}$, respectively.

Moreover, according to~\cite{related_work_Galiotto},
the area spectral efficiency (ASE) in $\textrm{bps/Hz/km}^{2}$ for a given $\lambda$ can be defined as
%\begin{singlespace}
\begin{equation}
A^{\textrm{ASE}}\left(\lambda,\gamma_{0}\right)=\lambda\int_{\gamma_{0}}^{\infty}\log_{2}\left(1+\gamma\right)f_{\mathit{\Gamma}}\left(\lambda,\gamma\right)d\gamma,\label{eq:ASE_def}
\end{equation}
%\end{singlespace}
\noindent where $\gamma_{0}$ is the minimum working SINR for the considered SCN,
and $f_{\mathit{\Gamma}}\left(\lambda,\gamma\right)$ is the probability density function (PDF) of SINR observed at the typical UE for a particular value of $\lambda$.
Note that the ASE defined in this paper is different from that in~\cite{related_work_Jeff},
where a deterministic rate based on $\gamma_{0}$ is assumed for the typical UE,
no matter what the actual SINR value is.
The ASE definition in (\ref{eq:ASE_def}) is more realistic due to the SINR-dependent rate,
but it is more complex to analyze, as it requires one more fold of numerical integral compared with~\cite{related_work_Jeff}.

Based on the definition of $p^{\textrm{cov}}\left(\lambda,\gamma\right)$ in (\ref{eq:Coverage_Prob_def}),
which is the complementary cumulative distribution function (CCDF) of SINR,
$f_{\mathit{\Gamma}}\left(\lambda,\gamma\right)$ can be computed by
%\begin{singlespace}
\begin{equation}
f_{\mathit{\Gamma}}\left(\lambda,\gamma\right)=\frac{\partial\left(1-p^{\textrm{cov}}\left(\lambda,\gamma\right)\right)}{\partial\gamma}.\label{eq:cond_SINR_PDF}
\end{equation}
%\end{singlespace}

Given the definition of the coverage probability and the ASE respectively presented in~(\ref{eq:Coverage_Prob_def}) and~(\ref{eq:ASE_def}),
in the following we will analyze the two performance measures for the considered UAS.
Based on the proposed path loss model in (\ref{eq:prop_PL_model}),
we present our main result on $p^{\textrm{cov}}\left(\lambda,\gamma\right)$ in Theorem~\ref{thm:p_cov_UAS1}.

\begin{thm}
\label{thm:p_cov_UAS1}Considering the path loss model in (\ref{eq:prop_PL_model}),
$p^{\textrm{cov}}\left(\lambda,\gamma\right)$ can be derived as
\begin{equation}
p^{\textrm{cov}}\left(\lambda,\gamma\right)=\sum_{n=1}^{N}\left(T_{n}^{\textrm{L}}+T_{n}^{\textrm{NL}}\right),\label{eq:Theorem_1_p_cov}
\end{equation}
where $T_{n}^{\textrm{L}}=\int_{d_{n-1}}^{d_{n}}\textrm{Pr}\left[\frac{P\zeta_{n}^{\textrm{L}}\left(r\right)h}{I_{r}+N_{0}}>\gamma\right]f_{R,n}^{\textrm{L}}\left(r\right)dr$,
$T_{n}^{\textrm{NL}}=\int_{d_{n-1}}^{d_{n}}\textrm{Pr}\left[\frac{P\zeta_{n}^{\textrm{NL}}\left(r\right)h}{I_{r}+N_{0}}>\gamma\right]f_{R,n}^{\textrm{NL}}\left(r\right)dr$,
and $d_{0}$ and $d_{N}$ are respectively defined as $0$ and $\infty$.
Moreover, $f_{R,n}^{\textrm{L}}\left(r\right)$ and $f_{R,n}^{\textrm{NL}}\left(r\right)$ are given by
\begin{eqnarray}
\hspace{-0.3cm}\hspace{-0.3cm}f_{R,n}^{\textrm{L}}\left(r\right)\hspace{-0.3cm} & = & \hspace{-0.3cm}\exp\left(-\int_{0}^{r_{1}}\left(1-\textrm{Pr}^{\textrm{L}}\left(u\right)\right)2\pi u\lambda du\right)\nonumber \\
\hspace{-0.3cm} &  & \hspace{-0.3cm}\times\exp\left(-\int_{0}^{r}\textrm{Pr}^{\textrm{L}}\left(u\right)2\pi u\lambda du\right)\nonumber \\
\hspace{-0.3cm} &  & \hspace{-0.3cm}\times\textrm{Pr}_{n}^{\textrm{L}}\left(r\right)\times2\pi r\lambda,\quad\left(d_{n-1}<r\leq d_{n}\right),\label{eq:geom_dis_PDF_UAS1_LoS_thm}
\end{eqnarray}
and
\begin{eqnarray}
\hspace{-0.3cm}\hspace{-0.3cm}f_{R,n}^{\textrm{NL}}\left(r\right)\hspace{-0.3cm} & = & \hspace{-0.3cm}\exp\left(-\int_{0}^{r_{2}}\textrm{Pr}^{\textrm{L}}\left(u\right)2\pi u\lambda du\right)\nonumber \\
\hspace{-0.3cm} &  & \hspace{-0.3cm}\times\exp\left(-\int_{0}^{r}\left(1-\textrm{Pr}^{\textrm{L}}\left(u\right)\right)2\pi u\lambda du\right)\nonumber \\
\hspace{-0.3cm} &  & \hspace{-0.3cm}\times\left(1-\textrm{Pr}_{n}^{\textrm{L}}\left(r\right)\right)\times2\pi r\lambda,\quad\left(d_{n-1}<r\leq d_{n}\right),\label{eq:geom_dis_PDF_UAS1_NLoS_thm}
\end{eqnarray}
where $r_{1}$ and $r_{2}$ are determined by
\begin{equation}
r_{1}=\underset{r_{1}}{\arg}\left\{ \zeta^{\textrm{NL}}\left(r_{1}\right)=\zeta_{n}^{\textrm{L}}\left(r\right)\right\} ,\label{eq:def_r_1}
\end{equation}
and
\begin{equation}
r_{2}=\underset{r_{2}}{\arg}\left\{ \zeta^{\textrm{L}}\left(r_{2}\right)=\zeta_{n}^{\textrm{NL}}\left(r\right)\right\} .\label{eq:def_r_2}
\end{equation}

Furthermore, $\textrm{Pr}\left[\frac{P\zeta_{n}^{\textrm{L}}\left(r\right)h}{I_{r}+N_{0}}>\gamma\right]$
and $\textrm{Pr}\left[\frac{P\zeta_{n}^{\textrm{NL}}\left(r\right)h}{I_{r}+N_{0}}>\gamma\right]$
are respectively computed by
\begin{equation}
\hspace{-0.1cm}\hspace{-0.1cm}\textrm{Pr}\left[\frac{P\zeta_{n}^{\textrm{L}}\left(r\right)h}{I_{r}+N_{0}}>\gamma\right]\hspace{-0.1cm}=\hspace{-0.1cm}\exp\left(\hspace{-0.1cm}-\frac{\gamma N_{0}}{P\zeta_{n}^{\textrm{L}}\left(r\right)}\right)\hspace{-0.1cm}\mathscr{L}_{I_{r}}\hspace{-0.1cm}\left(\frac{\gamma}{P\zeta_{n}^{\textrm{L}}\left(r\right)}\right),\label{eq:Pr_SINR_req_UAS1_LoS_thm}
\end{equation}
and
\begin{equation}
\textrm{Pr}\left[\frac{P\zeta_{n}^{\textrm{NL}}\left(r\right)h}{I_{r}+N_{0}}>\gamma\right]\hspace{-0.1cm}=\hspace{-0.1cm}\exp\left(\hspace{-0.1cm}-\frac{\gamma N_{0}}{P\zeta_{n}^{\textrm{NL}}\left(r\right)}\right)\hspace{-0.1cm}\mathscr{L}_{I_{r}}\hspace{-0.1cm}\left(\frac{\gamma}{P\zeta_{n}^{\textrm{NL}}\left(r\right)}\right),\label{eq:Pr_SINR_req_UAS1_NLoS_thm}
\end{equation}
where $\mathscr{L}_{I_{r}}\left(s\right)$ is the Laplace transform of $I_{r}$ evaluated at $s$.\end{thm}

\begin{IEEEproof}
See Appendix~A.
\end{IEEEproof}

\vspace{0.3cm}

As can be observed from Theorem~\ref{thm:p_cov_UAS1},
the piece-wise path loss function for LoS transmission $\left\{ \zeta_{n}^{\textrm{L}}\left(r\right)\right\} $,
the piece-wise path loss function for NLoS transmission $\left\{ \zeta_{n}^{\textrm{NL}}\left(r\right)\right\} $,
and the piece-wise LoS probability function $\left\{ \textrm{Pr}_{n}^{\textrm{L}}\left(r\right)\right\}$
play active roles in determining the final result of $p^{\textrm{cov}}\left(\lambda,\gamma\right)$.
We will investigate their impacts on network performance in detail in the following sections.

Plugging $p^{\textrm{cov}}\left(\lambda,\gamma\right)$ obtained from (\ref{eq:Theorem_1_p_cov}) into (\ref{eq:cond_SINR_PDF}),
we can get the result of the ASE using (\ref{eq:ASE_def}).

Regarding the computational process to obtain $p^{\textrm{cov}}\left(\lambda,\gamma\right)$ presented in Theorem~\ref{thm:p_cov_UAS1},
for a general case,
three folds of integrals are respectively required for the calculation of
$\left\{ f_{R,n}^{Path}\left(r\right)\right\} $, $\left\{ \mathscr{L}_{I_{r}}\left(\frac{\gamma}{P\zeta_{n}^{Path}\left(r\right)}\right)\right\} $ and $\left\{ T_{n}^{Path}\right\} $,
where the string variable $Path$ takes the value of ``L'' (for the LoS case) or ``NL'' (for the NLoS case).
Note that an additional fold of integral is needed in (\ref{eq:ASE_def}) for the calculation of $A^{\textrm{ASE}}\left(\lambda,\gamma_{0}\right)$,
making it a 4-fold integral computation.

\section{Study of a 3GPP Special Case \label{sec:A-3GPP-Special-Case}}

As a special case for Theorem~\ref{thm:p_cov_UAS1},
we consider the following path loss function, $\zeta\left(r\right)$,
adopted by the 3GPP~\cite{TR36.828}, i.e.,
\begin{equation}
\hspace{-0.2cm}\zeta\left(r\right)=\begin{cases}
\begin{array}{l}
A^{{\rm {L}}}r^{-\alpha^{{\rm {L}}}},\\
A^{{\rm {NL}}}r^{-\alpha^{{\rm {NL}}}},
\end{array}\hspace{-0.3cm}\hspace{-0.3cm} & \begin{array}{l}
\textrm{with probability }\textrm{Pr}^{\textrm{L}}\left(r\right)\\
\textrm{with probability }\left(1-\textrm{Pr}^{\textrm{L}}\left(r\right)\right)
\end{array}\end{cases}\hspace{-0.2cm},\label{eq:PL_BS2UE_2slopes}
\end{equation}
together with a linear LoS probability function of $\textrm{Pr}^{\textrm{L}}\left(r\right)$,
also adopted by the 3GPP~\cite{SCM_pathloss_model}, i.e.,
%\begin{singlespace}
\begin{equation}
\textrm{Pr}^{\textrm{L}}\left(r\right)=\begin{cases}
\begin{array}{l}
1-\frac{r}{d_{1}},\\
0,
\end{array}\hspace{-0.3cm} & \begin{array}{l}
0<r\leq d_{1}\\
r>d_{1}
\end{array}\end{cases},\label{eq:LoS_Prob_func_linear}
\end{equation}
%\end{singlespace}
where $d_{1}$ is a parameter that determines the decreasing slope of the linear function $\textrm{Pr}^{\textrm{L}}\left(r\right)$.

Considering the general path loss model presented in (\ref{eq:prop_PL_model}),
the path loss model shown in (\ref{eq:PL_BS2UE_2slopes}) and (\ref{eq:LoS_Prob_func_linear}) can be deemed as a special case of (\ref{eq:prop_PL_model}) with the following substitution:
$N=2$, $\zeta_{1}^{\textrm{L}}\left(r\right)=\zeta_{2}^{\textrm{L}}\left(r\right)=A^{{\rm {L}}}r^{-\alpha^{{\rm {L}}}}$,
$\zeta_{1}^{\textrm{NL}}\left(r\right)=\zeta_{2}^{\textrm{NL}}\left(r\right)=A^{{\rm {NL}}}r^{-\alpha^{{\rm {NL}}}}$,
$\textrm{Pr}_{1}^{\textrm{L}}\left(r\right)=1-\frac{r}{d_{1}}$, and
$\textrm{Pr}_{2}^{\textrm{L}}\left(r\right)=0$.
For clarity, this 3GPP special case is referred to as 3GPP Case~1 in the sequel.

It should be noted that 3GPP Case 1 is compatible with dense SCNs,
because both the exponential path loss function in~\cite{TR36.828} and the LoS probability function in~\cite{SCM_pathloss_model}  are valid for small cells only
(referred to as microcells in~\cite{SCM_pathloss_model}).
It should also be noted that there is another LoS probability function defined in~\cite{TR36.828},
which takes a more complicated mathematical form given by
%\begin{singlespace}
%\noindent
\begin{eqnarray}
\textrm{Pr}^{\textrm{L}}\left(r\right)\hspace{-0.3cm} & = & \hspace{-0.3cm}0.5-\min\left\{ {0.5,5\exp\left({-\frac{R_{1}}{r}}\right)}\right\} \nonumber \\
\hspace{-0.3cm} &  & \hspace{-0.3cm}+\min\left\{ {0.5,5\exp\left({-\frac{r}{R_{2}}}\right)}\right\} ,\label{eq:LoS_Prob_func_EXP}
\end{eqnarray}
%\end{singlespace}
where $R_{1}$ and $R_{2}$ are shape parameters to ensure the continuity of $\textrm{Pr}^{\textrm{L}}\left(r\right)$.

To show how $\textrm{Pr}^{\textrm{L}}\left(r\right)$ in (\ref{eq:LoS_Prob_func_EXP}) can be fitted into our proposed path loss model,
we can re-formulate (\ref{eq:LoS_Prob_func_EXP}) as
%\begin{singlespace}
%\noindent
\begin{equation}
\textrm{Pr}^{\textrm{L}}\left(r\right)=\begin{cases}
\begin{array}{l}
1-5\exp\left(-R_{1}/r\right),\\
5\exp\left(-r/R_{2}\right),
\end{array} & \begin{array}{l}
0<r\leq d_{1}\\
r>d_{1}
\end{array}\end{cases},\label{eq:LoS_Prob_func_reverseS_shape}
\end{equation}
%\end{singlespace}
where $d_{1}=\frac{R_{1}}{\ln10}$.

The combination of the LoS probability function in (\ref{eq:LoS_Prob_func_reverseS_shape}) and the path loss function in (\ref{eq:PL_BS2UE_2slopes}) can then be deemed as a special case of the proposed path loss model in (\ref{eq:prop_PL_model}) with the following substitution: $N=2$, $\zeta_{1}^{\textrm{L}}\left(r\right)=\zeta_{2}^{\textrm{L}}\left(r\right)=A^{{\rm {L}}}r^{-\alpha^{{\rm {L}}}}$,
$\zeta_{1}^{\textrm{NL}}\left(r\right)=\zeta_{2}^{\textrm{NL}}\left(r\right)=A^{{\rm {NL}}}r^{-\alpha^{{\rm {NL}}}}$,
$\textrm{Pr}_{1}^{\textrm{L}}\left(r\right)=1-5\exp\left(-R_{1}/r\right)$,
and $\textrm{Pr}_{2}^{\textrm{L}}\left(r\right)=5\exp\left(-r/R_{2}\right)$.
For clarity, this combined case with both the path loss function and the LoS probability function coming from~\cite{TR36.828} is referred to as 3GPP Case~2 in our paper.
Note that 3GPP Case~2 was treated in~\cite{related_work_Galiotto} by approximating $\textrm{Pr}^{\textrm{L}}\left(r\right)$ in (\ref{eq:LoS_Prob_func_EXP}) as an exponentially decreasing function.
However, the results in~\cite{related_work_Galiotto} are less tractable
than ours as well as those in~\cite{related_work_Jeff} and~\cite{Related_work_Health},
because the approximated function is still difficult to deal with in the theoretical analysis.

In the following, we first investigate 3GPP Case~1 in our case study,
because it gives more tractable results than 3GPP Case~2 treated in~\cite{related_work_Galiotto},
as will be shown in the following subsections.
Thereafter, we will numerically investigate 3GPP Case~2 using Theorem~\ref{thm:p_cov_UAS1} in Section~\ref{sec:Simulation-and-Discussion},
and will show that similar conclusions like those for 3GPP Case~1 can also be drawn for 3GPP Case~2.
More importantly, we will extend 3GPP Case~1 to study an approximated 3GPP Case~2 in Section~\ref{sec:Simulation-and-Discussion},
thus showing the usefulness of studying 3GPP Case~1 with the \emph{linear} LoS probability function.

To sum up, taking the linear LoS probability function from~\cite{SCM_pathloss_model} to create 3GPP Case~1 not only allows us to obtain more tractable results,
but also help us to deal with more complicated path loss models in practice.

For 3GPP Case~1, according to Theorem~\ref{thm:p_cov_UAS1},
$p^{\textrm{cov}}\left(\lambda,\gamma\right)$ can then be computed by
%\noindent
\begin{equation}
p^{\textrm{cov}}\left(\lambda,\gamma\right)=\sum_{n=1}^{2}\left(T_{n}^{\textrm{L}}+T_{n}^{\textrm{NL}}\right).\label{eq:p_cov_special_case_UAS1}
\end{equation}

In the following subsections,
we investigate the computation of $T_{1}^{\textrm{L}}$, $T_{1}^{\textrm{NL}}$, $T_{2}^{\textrm{L}}$, and $T_{2}^{\textrm{NL}}$,
respectively.

\subsection{The Computation of $T_{1}^{\textrm{L}}$\label{sub:The-Computation-of-T1L}}

From Theorem~\ref{thm:p_cov_UAS1},
$T_{1}^{\textrm{L}}$ can be obtained as
\begin{eqnarray}
\hspace{-0.2cm}T_{1}^{\textrm{L}}\hspace{-0.3cm} & = & \hspace{-0.3cm}\int_{0}^{d_{1}}\hspace{-0.1cm}\exp\left(-\frac{\gamma N_{0}}{P\zeta_{1}^{\textrm{L}}\left(r\right)}\right)\hspace{-0.1cm}\mathscr{L}_{I_{r}}\hspace{-0.1cm}\left(\frac{\gamma}{P\zeta_{1}^{\textrm{L}}\left(r\right)}\right)\hspace{-0.1cm}f_{R,1}^{\textrm{L}}\left(r\right)dr\nonumber \\
\hspace{-0.3cm} & \overset{(a)}{=} & \hspace{-0.3cm}\int_{0}^{d_{1}}\hspace{-0.1cm}\exp\left(-\frac{\gamma r^{\alpha^{{\rm {L}}}}N_{0}}{PA^{{\rm {L}}}}\right)\hspace{-0.1cm}\mathscr{L}_{I_{r}}\hspace{-0.1cm}\left(\frac{\gamma r^{\alpha^{{\rm {L}}}}}{PA^{{\rm {L}}}}\right)\hspace{-0.1cm}f_{R,1}^{\textrm{L}}\left(r\right)dr,\label{eq:T_1_UAS1_LoS}
\end{eqnarray}
where $\zeta_{1}^{\textrm{L}}\left(r\right)=A^{{\rm {L}}}r^{-\alpha^{{\rm {L}}}}$ from (\ref{eq:PL_BS2UE_2slopes}) is plugged into the step (a) of (\ref{eq:T_1_UAS1_LoS})
 and $\mathscr{L}_{I_{r}}\left(s\right)$ is the Laplace transform of RV $I_{r}$ evaluated at $s$.

In (\ref{eq:T_1_UAS1_LoS}),
according to Theorem~\ref{thm:p_cov_UAS1} and (\ref{eq:LoS_Prob_func_linear}), we have
\begin{eqnarray}
f_{R,1}^{\textrm{L}}\left(r\right)\hspace{-0.3cm} & = & \hspace{-0.3cm}\exp\left(-\int_{0}^{r_{1}}\lambda\frac{u}{d_{1}}2\pi udu\right)\nonumber \\
\hspace{-0.3cm} &  & \hspace{-0.3cm}\times\exp\left(-\int_{0}^{r}\hspace{-0.1cm}\lambda\left(1-\frac{u}{d_{1}}\right)2\pi udu\right)\hspace{-0.1cm}\left(1-\frac{r}{d_{1}}\right)\hspace{-0.1cm}2\pi r\lambda\nonumber \\
\hspace{-0.3cm} & = & \hspace{-0.3cm}\exp\left(-\pi\lambda r^{2}+2\pi\lambda\left(\frac{r^{3}}{3d_{1}}-\frac{r_{1}^{3}}{3d_{1}}\right)\right)\nonumber \\
\hspace{-0.3cm} &  & \hspace{-0.3cm}\times\left(1-\frac{r}{d_{1}}\right)2\pi r\lambda,\quad\left(0<r\leq d_{1}\right),\label{eq:spec_geom_dis_PDF_UAS1_LoS_seg1}
\end{eqnarray}
where $r_{1}=\left(\frac{A^{{\rm {NL}}}}{A^{{\rm {L}}}}\right)^{\frac{1}{\alpha^{{\rm {NL}}}}}r^{\frac{\alpha^{{\rm {L}}}}{\alpha^{{\rm {NL}}}}}$
according to~(\ref{eq:def_r_1}).

Besides, to compute $\mathscr{L}_{I_{r}}\left(\frac{\gamma r^{\alpha^{{\rm {L}}}}}{PA^{{\rm {L}}}}\right)$ in (\ref{eq:T_1_UAS1_LoS}) for the range of $0<r\leq d_{1}$,
we propose Lemma~\ref{lem:laplace_term_UAS1_LoS_seg1}.
\begin{lem}
\noindent \label{lem:laplace_term_UAS1_LoS_seg1}$\mathscr{L}_{I_{r}}\left(\frac{\gamma r^{\alpha^{{\rm {L}}}}}{PA^{{\rm {L}}}}\right)$
in the range of $0<r\leq d_{1}$ can be calculated by
\noindent
\begin{eqnarray}
\hspace{-0.4cm} & \hspace{-0.3cm}\hspace{-0.1cm} & \mathscr{L}_{I_{r}}\left(\frac{\gamma r^{\alpha^{{\rm {L}}}}}{PA^{{\rm {L}}}}\right)=\nonumber \\
\hspace{-0.4cm} & \hspace{-0.3cm} & \exp\hspace{-0.1cm}\left(\hspace{-0.1cm}-2\pi\lambda\hspace{-0.1cm}\left(\hspace{-0.1cm}\rho_{1}\hspace{-0.1cm}\left(\hspace{-0.1cm}\alpha^{{\rm {L}}},1,\left(\gamma r^{\alpha^{{\rm {L}}}}\right)^{\hspace{-0.1cm}-1}\hspace{-0.1cm}\hspace{-0.1cm},d_{1}\hspace{-0.1cm}\right)\hspace{-0.1cm}-\hspace{-0.1cm}\rho_{1}\hspace{-0.1cm}\left(\hspace{-0.1cm}\alpha^{{\rm {L}}},1,\left(\gamma r^{\alpha^{{\rm {L}}}}\right)^{\hspace{-0.1cm}-1}\hspace{-0.1cm}\hspace{-0.1cm},r\hspace{-0.1cm}\right)\hspace{-0.1cm}\right)\hspace{-0.1cm}\right)\nonumber \\
\hspace{-0.4cm} & \hspace{-0.3cm} & \times\exp\hspace{-0.1cm}\left(\frac{2\pi\lambda}{d_{1}}\hspace{-0.1cm}\left(\hspace{-0.1cm}\rho_{1}\hspace{-0.1cm}\left(\hspace{-0.1cm}\alpha^{{\rm {L}}},2,\left(\gamma r^{\alpha^{{\rm {L}}}}\right)^{\hspace{-0.1cm}-1}\hspace{-0.1cm}\hspace{-0.1cm},d_{1}\hspace{-0.1cm}\right)\hspace{-0.1cm}-\hspace{-0.1cm}\rho_{1}\hspace{-0.1cm}\left(\hspace{-0.1cm}\alpha^{{\rm {L}}},2,\left(\gamma r^{\alpha^{{\rm {L}}}}\right)^{\hspace{-0.1cm}-1}\hspace{-0.1cm}\hspace{-0.1cm},r\hspace{-0.1cm}\right)\hspace{-0.1cm}\right)\hspace{-0.1cm}\right)\nonumber \\
\hspace{-0.4cm} & \hspace{-0.3cm} & \times\exp\left(-\frac{2\pi\lambda}{d_{1}}\rho_{1}\left(\alpha^{{\rm {NL}}},2,\left(\frac{\gamma A^{{\rm {NL}}}}{A^{{\rm {L}}}}r^{\alpha^{{\rm {L}}}}\right)^{-1}\hspace{-0.1cm}\hspace{-0.1cm},d_{1}\right)\right.\nonumber \\
\hspace{-0.4cm} & \hspace{-0.3cm} & \qquad\qquad\left.+\frac{2\pi\lambda}{d_{1}}\rho_{1}\left(\alpha^{{\rm {NL}}},2,\left(\frac{\gamma A^{{\rm {NL}}}}{A^{{\rm {L}}}}r^{\alpha^{{\rm {L}}}}\right)^{-1}\hspace{-0.1cm}\hspace{-0.1cm},r_{1}\right)\right)\nonumber \\
\hspace{-0.4cm} & \hspace{-0.3cm} & \times\exp\hspace{-0.1cm}\left(\hspace{-0.1cm}-2\pi\lambda\rho_{2}\hspace{-0.1cm}\left(\hspace{-0.1cm}\alpha^{{\rm {NL}}},1,\left(\frac{\gamma A^{{\rm {NL}}}}{A^{{\rm {L}}}}r^{\alpha^{{\rm {L}}}}\right)^{\hspace{-0.1cm}-1}\hspace{-0.1cm}\hspace{-0.1cm}\hspace{-0.1cm},d_{1}\hspace{-0.1cm}\right)\hspace{-0.1cm}\right)\hspace{-0.1cm},\left(0<r\leq d_{1}\right)\nonumber \\
\hspace{-0.4cm} & \hspace{-0.3cm}\label{eq:Lemma_3}
\end{eqnarray}
\vspace{-0.3cm}
\hspace{-0.3cm} where
\begin{equation}
\rho_{1}\left(\alpha,\beta,t,d\right)=\left[\frac{d^{\left(\beta+1\right)}}{\beta+1}\right]{}_{2}F_{1}\left[1,\frac{\beta+1}{\alpha};1+\frac{\beta+1}{\alpha};-td^{\alpha}\right],\label{eq:rou1_func}
\end{equation}
and
\begin{equation}
\hspace{-4.3cm}\rho_{2}\left(\alpha,\beta,t,d\right)=\left[\frac{d^{-\left(\alpha-\beta-1\right)}}{t\left(\alpha-\beta-1\right)}\right]\nonumber
\end{equation}
\begin{equation}
\hspace{-0.3cm}\times{}_{2}F_{1}\left[1,1-\frac{\beta+1}{\alpha};2-\frac{\beta+1}{\alpha};-\frac{1}{td^{\alpha}}\right],\left(\alpha>\beta+1\right),\label{eq:rou2_func}
\end{equation}
where $_{2}F_{1}\left[\cdot,\cdot;\cdot;\cdot\right]$ is the hyper-geometric function~\cite{Book_Integrals}.
\end{lem}

\begin{IEEEproof}
See Appendix~B.
\end{IEEEproof}

\vspace{0.3cm}

To sum up, $T_{1}^{\textrm{L}}$ can be evaluated as
\begin{equation}
\hspace{-0.3cm}T_{1}^{\textrm{L}}=\int_{0}^{d_{1}}\exp\left(-\frac{\gamma r^{\alpha^{{\rm {L}}}}N_{0}}{PA^{{\rm {L}}}}\right)\mathscr{L}_{I_{r}}\left(\frac{\gamma r^{\alpha^{{\rm {L}}}}}{PA^{{\rm {L}}}}\right)f_{R,1}^{\textrm{L}}\left(r\right)dr,\label{eq:T_1_UAS1_LoS_final}
\end{equation}
where $f_{R,1}^{\textrm{L}}\left(r\right)$ and $\mathscr{L}_{I_{r}}\left(\frac{\gamma r^{\alpha^{{\rm {L}}}}}{PA^{{\rm {L}}}}\right)$
are given by (\ref{eq:spec_geom_dis_PDF_UAS1_LoS_seg1}) and (\ref{eq:Lemma_3}),
respectively.

\subsection{The Computation of $T_{1}^{\textrm{NL}}$\label{sub:The-Computation-of-T1NL}}

From Theorem~\ref{thm:p_cov_UAS1},
$T_{1}^{\textrm{NL}}$ can be obtained as
\begin{eqnarray}
\hspace{-0.2cm}T_{1}^{\textrm{NL}}\hspace{-0.3cm} & = & \hspace{-0.3cm}\int_{0}^{d_{1}}\exp\left(-\frac{\gamma N_{0}}{P\zeta_{1}^{\textrm{NL}}\left(r\right)}\right)\mathscr{L}_{I_{r}}\left(\frac{\gamma}{P\zeta_{1}^{\textrm{NL}}\left(r\right)}\right)f_{R,1}^{\textrm{NL}}\left(r\right)dr\nonumber \\
\hspace{-0.2cm}\hspace{-0.3cm} & \overset{(a)}{=} & \hspace{-0.3cm}\int_{0}^{d_{1}}\exp\left(-\frac{\gamma r^{\alpha^{{\rm {NL}}}}N_{0}}{PA^{{\rm {NL}}}}\right)\mathscr{L}_{I_{r}}\left(\frac{\gamma r^{\alpha^{{\rm {NL}}}}}{PA^{{\rm {NL}}}}\right)f_{R,1}^{\textrm{NL}}\left(r\right)dr,\nonumber\\
\label{eq:T_1_UAS1_NLoS}
\end{eqnarray}
where $\zeta_{1}^{\textrm{NL}}\left(r\right)=A^{{\rm {NL}}}r^{-\alpha^{{\rm {NL}}}}$ from (\ref{eq:PL_BS2UE_2slopes}) is plugged into the step (a) of (\ref{eq:T_1_UAS1_NLoS}).

In (\ref{eq:T_1_UAS1_NLoS}),
according to Theorem~\ref{thm:p_cov_UAS1} and (\ref{eq:LoS_Prob_func_linear}), we have
\begin{eqnarray}
f_{R,1}^{\textrm{NL}}\left(r\right)\hspace{-0.3cm} & = & \hspace{-0.3cm}\exp\left(-\int_{0}^{r_{2}}\lambda\textrm{Pr}^{\textrm{L}}\left(u\right)2\pi udu\right)\nonumber \\
\hspace{-0.3cm} &  & \hspace{-0.3cm}\times\exp\left(-\int_{0}^{r}\lambda\left(1-\textrm{Pr}^{\textrm{L}}\left(u\right)\right)2\pi udu\right)\nonumber \\
\hspace{-0.3cm} &  & \hspace{-0.3cm}\times\left(\frac{r}{d_{1}}\right)2\pi r\lambda,\quad\left(0<r\leq d_{1}\right),\label{eq:spec_geom_dis_PDF_UAS1_NLoS_seg1}
\end{eqnarray}
where $r_{2}=\left(\frac{A^{{\rm {L}}}}{A^{{\rm {NL}}}}\right)^{\frac{1}{\alpha^{{\rm {L}}}}}r^{\frac{\alpha^{{\rm {NL}}}}{\alpha^{{\rm {L}}}}}$ according to~(\ref{eq:def_r_2}).
Since the numerical relationship between $r_{2}$ and $d_{1}$ affects the calculation of the first multiplier in (\ref{eq:spec_geom_dis_PDF_UAS1_NLoS_seg1}),
i.e., $\exp\left(-\int_{0}^{r_{2}}\lambda\textrm{Pr}^{\textrm{L}}\left(u\right)2\pi udu\right)$,
we should discuss the cases of $0<r_{2}\leq d_{1}$ and $r_{2}>d_{1}$.

If $0<r_{2}\leq d_{1}$, i.e., $0<r\leq y_{1}=d_{1}^{\frac{\alpha^{{\rm {L}}}}{\alpha^{{\rm {NL}}}}}\left(\frac{A^{{\rm {NL}}}}{A^{{\rm {L}}}}\right)^{\frac{1}{\alpha^{{\rm {NL}}}}}$,
we have
\begin{eqnarray}
f_{R,1}^{\textrm{NL}}\left(r\right)\hspace{-0.3cm} & = & \hspace{-0.3cm}\exp\left(-\int_{0}^{r_{2}}\lambda\left(1-\frac{u}{d_{1}}\right)2\pi udu\right)\nonumber \\
\hspace{-0.3cm} &   & \hspace{-0.3cm}\times\exp\left(-\int_{0}^{r}\lambda\frac{u}{d_{1}}2\pi udu\right)\left(\frac{r}{d_{1}}\right)2\pi r\lambda\nonumber \\
\hspace{-0.3cm} & = & \hspace{-0.3cm}\exp\left(-\pi\lambda r_{2}^{2}+2\pi\lambda\left(\frac{r_{2}^{3}}{3d_{1}}-\frac{r^{3}}{3d_{1}}\right)\right)\nonumber \\
\hspace{-0.3cm} &   & \hspace{-0.3cm}\times\left(\frac{r}{d_{1}}\right)2\pi r\lambda,\quad\left(0<r\leq y_{1}\right).\label{eq:spec_geom_dis_PDF_UAS1_NLoS_seg1_case1}
\end{eqnarray}
Otherwise, if $r_{2}>d_{1}$, i.e., $y_{1}<r\leq d_{1}$,
we have
\begin{eqnarray}
f_{R,1}^{\textrm{NL}}\left(r\right)\hspace{-0.3cm} & = & \hspace{-0.3cm}\exp\left(-\int_{0}^{d_{1}}\lambda\left(1-\frac{u}{d_{1}}\right)2\pi udu\right)\nonumber \\
\hspace{-0.3cm} &   & \hspace{-0.3cm}\times\exp\left(-\int_{0}^{r}\lambda\frac{u}{d_{1}}2\pi udu\right)\left(\frac{r}{d_{1}}\right)2\pi r\lambda\nonumber \\
\hspace{-0.3cm} & = & \hspace{-0.3cm}\exp\left(-\frac{\pi\lambda d_{1}^{2}}{3}-\frac{2\pi\lambda r^{3}}{3d_{1}}\right)\nonumber \\
\hspace{-0.3cm} &   & \hspace{-0.3cm}\times\left(\frac{r}{d_{1}}\right)2\pi r\lambda,\quad\left(y_{1}<r\leq d_{1}\right).\label{eq:spec_geom_dis_PDF_UAS1_NLoS_seg1_case2}
\end{eqnarray}

Besides, to compute $\mathscr{L}_{I_{r}}\left(\frac{\gamma r^{\alpha^{{\rm {NL}}}}}{PA^{{\rm {NL}}}}\right)$ in (\ref{eq:T_1_UAS1_NLoS}) for the range of $0<r\leq d_{1}$,
we propose Lemma~\ref{lem:laplace_term_UAS1_NLoS_seg1} in the following.
Note that since the calculation of $f_{R,1}^{\textrm{NL}}\left(r\right)$ is divided into two cases respectively shown in (\ref{eq:spec_geom_dis_PDF_UAS1_NLoS_seg1_case1}) and (\ref{eq:spec_geom_dis_PDF_UAS1_NLoS_seg1_case2}),
the calculation of $\mathscr{L}_{I_{r}}\left(\frac{\gamma r^{\alpha^{{\rm {NL}}}}}{PA^{{\rm {NL}}}}\right)$
%in the range of $0<r\leq d_{1}$
should also be divided into those two cases,
%because the interference is integrated from distance $r$ to infinity.
because the first case includes both LoS and NLoS interference, while the second case only considers NLoS interference.

\begin{lem}
\noindent \label{lem:laplace_term_UAS1_NLoS_seg1}$\mathscr{L}_{I_{r}}\left(\frac{\gamma r^{\alpha^{{\rm {NL}}}}}{PA^{{\rm {NL}}}}\right)$ in the range of $0<r\leq d_{1}$ can be divided into two cases, i.e., $0<r\leq y_{1}$ and $y_{1}<r\leq d_{1}$.
The results are as follows,
\begin{eqnarray}
\hspace{-1cm} & \hspace{-0.3cm} & \hspace{-0.6cm}\mathscr{L}_{I_{r}}\left(\frac{\gamma r^{\alpha^{{\rm {NL}}}}}{PA^{{\rm {NL}}}}\right)=\nonumber \\
 &  & \:\;\exp\left(-2\pi\lambda\rho_{1}\left(\alpha^{{\rm {L}}},1,\left(\frac{\gamma A^{{\rm {L}}}}{A^{{\rm {NL}}}}r^{\alpha^{{\rm {NL}}}}\right)^{-1},d_{1}\right)\right.\nonumber \\
 &  & \:\;\qquad\qquad\left.+2\pi\lambda\rho_{1}\left(\alpha^{{\rm {L}}},1,\left(\frac{\gamma A^{{\rm {L}}}}{A^{{\rm {NL}}}}r^{\alpha^{{\rm {NL}}}}\right)^{-1},r_{2}\right)\right)\nonumber \\
 &  & \:\;\times\exp\left(\frac{2\pi\lambda}{d_{1}}\rho_{1}\left(\alpha^{{\rm {L}}},2,\left(\frac{\gamma A^{{\rm {L}}}}{A^{{\rm {NL}}}}r^{\alpha^{{\rm {NL}}}}\right)^{-1},d_{1}\right)\right.\nonumber \\
 &  & \:\;\qquad\qquad\left.-\frac{2\pi\lambda}{d_{1}}\rho_{1}\left(\alpha^{{\rm {L}}},2,\left(\frac{\gamma A^{{\rm {L}}}}{A^{{\rm {NL}}}}r^{\alpha^{{\rm {NL}}}}\right)^{-1},r_{2}\right)\right)\nonumber \\
 &  & \:\;\times\exp\left(-\frac{2\pi\lambda}{d_{1}}\rho_{1}\left(\alpha^{{\rm {NL}}},2,\left(\gamma r^{\alpha^{{\rm {NL}}}}\right)^{-1},d_{1}\right)\right.\nonumber \\
 &  & \:\;\qquad\qquad\left.+\frac{2\pi\lambda}{d_{1}}\rho_{1}\left(\alpha^{{\rm {NL}}},2,\left(\gamma r^{\alpha^{{\rm {NL}}}}\right)^{-1},r\right)\right)\nonumber \\
 &  & \:\;\times\exp\left(-2\pi\lambda\rho_{2}\left(\alpha^{{\rm {NL}}},1,\left(\gamma r^{\alpha^{{\rm {NL}}}}\right)^{-1},d_{1}\right)\right),\nonumber \\
 &  & \:\;\hspace{5cm}\left(0<r\leq y_{1}\right),\label{eq:Lemma_4-1}
\end{eqnarray}
and
\begin{eqnarray}
\hspace{-1cm} & \hspace{-0.3cm} & \hspace{-0.6cm}\mathscr{L}_{I_{r}}\left(\frac{\gamma r^{\alpha^{{\rm {NL}}}}}{PA^{{\rm {NL}}}}\right)=\nonumber \\
 &  & \:\;\exp\left(-\frac{2\pi\lambda}{d_{1}}\rho_{1}\left(\alpha^{{\rm {NL}}},2,\left(\gamma r^{\alpha^{{\rm {NL}}}}\right)^{-1},d_{1}\right)\right.\nonumber \\
 &  & \:\;\qquad\qquad\left.+\frac{2\pi\lambda}{d_{1}}\rho_{1}\left(\alpha^{{\rm {NL}}},2,\left(\gamma r^{\alpha^{{\rm {NL}}}}\right)^{-1},r\right)\right)\nonumber \\
 &  & \:\;\times\exp\left(-2\pi\lambda\rho_{2}\left(\alpha^{{\rm {NL}}},1,\left(\gamma r^{\alpha^{{\rm {NL}}}}\right)^{-1},d_{1}\right)\right),\nonumber \\
 &  & \:\;\hspace{4.5cm}\left(y_{1}<r\leq d_{1}\right),\label{eq:Lemma_4-2}
\end{eqnarray}
where $\rho_{1}\left(\alpha,\beta,t,d\right)$ and $\rho_{2}\left(\alpha,\beta,t,d\right)$ are defined in (\ref{eq:rou1_func}) and (\ref{eq:rou2_func}), respectively.
\end{lem}

\begin{IEEEproof}
See Appendix~C.
\end{IEEEproof}

To sum up, $T_{1}^{\textrm{NL}}$ can be evaluated as
%\begin{eqnarray}
%T_{1}^{\textrm{NL}}\hspace{-0.3cm} & = & \hspace{-0.3cm}\int_{0}^{y_{1}}\exp\left(\hspace{-0.1cm}-\frac{\gamma r^{\alpha^{{\rm {NL}}}}N_{0}}{PA^{{\rm {NL}}}}\hspace{-0.1cm}\right)dr\nonumber \\
%\hspace{-0.3cm} &  & \hspace{-0.3cm}\qquad\times\left[\hspace{-0.1cm}\left.\mathscr{L}_{I_{r}}\hspace{-0.1cm}\left(\hspace{-0.1cm}\frac{\gamma r^{\alpha^{{\rm {NL}}}}}{PA^{{\rm {NL}}}}\hspace{-0.1cm}\right)\hspace{-0.1cm}f_{R,1}^{\textrm{NL}}\hspace{-0.1cm}\left(r\right)\right|0<r\leq y_{1}\hspace{-0.1cm}\right]\hspace{-0.1cm}dr\nonumber \\
%\hspace{-0.3cm} &  & \hspace{-0.3cm}+\int_{y_{1}}^{d_{1}}\exp\left(\hspace{-0.1cm}-\frac{\gamma r^{\alpha^{{\rm {NL}}}}N_{0}}{PA^{{\rm {NL}}}}\hspace{-0.1cm}\right)\nonumber \\
%\hspace{-0.3cm} &  & \hspace{-0.3cm}\qquad\times\left[\hspace{-0.1cm}\left.\mathscr{L}_{I_{r}}\hspace{-0.1cm}\left(\hspace{-0.1cm}\frac{\gamma r^{\alpha^{{\rm {NL}}}}}{PA^{{\rm {NL}}}}\hspace{-0.1cm}\right)\hspace{-0.1cm}f_{R,1}^{\textrm{NL}}\hspace{-0.1cm}\left(r\right)\right|y_{1}<r\leq d_{1}\hspace{-0.1cm}\right]\hspace{-0.1cm}dr,\label{eq:T_1_UAS1_NLoS_final}
%\end{eqnarray}
\begin{eqnarray}
T_{1}^{\textrm{NL}}\hspace{-0.3cm} & = & \hspace{-0.3cm}\int_{0}^{y_{1}}\exp\left(\hspace{-0.1cm}-\frac{\gamma r^{\alpha^{{\rm {NL}}}}N_{0}}{PA^{{\rm {NL}}}}\hspace{-0.1cm}\right)\nonumber \\
\hspace{-0.3cm} &  & \hspace{-0.3cm}\qquad\times\left[\hspace{-0.1cm}\left.\mathscr{L}_{I_{r}}\hspace{-0.1cm}\left(\hspace{-0.1cm}\frac{\gamma r^{\alpha^{{\rm {NL}}}}}{PA^{{\rm {NL}}}}\hspace{-0.1cm}\right)\hspace{-0.1cm}f_{R,1}^{\textrm{NL}}\hspace{-0.1cm}\left(r\right)\right|0<r\leq y_{1}\hspace{-0.1cm}\right]\hspace{-0.1cm}dr\nonumber
\end{eqnarray}
\begin{eqnarray}
\hspace{-0.3cm} &  & \hspace{-0.3cm}+\int_{y_{1}}^{d_{1}}\exp\left(\hspace{-0.1cm}-\frac{\gamma r^{\alpha^{{\rm {NL}}}}N_{0}}{PA^{{\rm {NL}}}}\hspace{-0.1cm}\right)\nonumber \\
\hspace{-0.3cm} &  & \hspace{-0.3cm}\qquad\times\left[\hspace{-0.1cm}\left.\mathscr{L}_{I_{r}}\hspace{-0.1cm}\left(\hspace{-0.1cm}\frac{\gamma r^{\alpha^{{\rm {NL}}}}}{PA^{{\rm {NL}}}}\hspace{-0.1cm}\right)\hspace{-0.1cm}f_{R,1}^{\textrm{NL}}\hspace{-0.1cm}\left(r\right)\right|y_{1}<r\leq d_{1}\hspace{-0.1cm}\right]\hspace{-0.1cm}dr,\label{eq:T_1_UAS1_NLoS_final}
\end{eqnarray}
%where $f_{R,1}^{\textrm{NL}}\left(r\right)$ is computed using (\ref{eq:spec_geom_dis_PDF_UAS1_NLoS_seg1_case1}) and (\ref{eq:spec_geom_dis_PDF_UAS1_NLoS_seg1_case2}),
%and $\mathscr{L}_{I_{r}}\left(\frac{\gamma r^{\alpha^{{\rm {NL}}}}}{PA^{{\rm {NL}}}}\right)$ is given by (\ref{eq:Lemma_4-1}) and (\ref{eq:Lemma_4-2}).
where (\ref{eq:spec_geom_dis_PDF_UAS1_NLoS_seg1_case1}), (\ref{eq:spec_geom_dis_PDF_UAS1_NLoS_seg1_case2}), (\ref{eq:Lemma_4-1}) and (\ref{eq:Lemma_4-2}) are plugged into (\ref{eq:T_1_UAS1_NLoS_final}).

\subsection{The Computation of $T_{2}^{\textrm{L}}$\label{sub:The-Computation-of-T2L}}

From Theorem~\ref{thm:p_cov_UAS1},
$T_{2}^{\textrm{L}}$ can be derived as
\begin{equation}
\hspace{-0.3cm}T_{2}^{\textrm{L}}=\int_{d_{1}}^{\infty}\hspace{-0.1cm}\exp\left(\hspace{-0.1cm}-\frac{\gamma N_{0}}{P\zeta_{2}^{\textrm{L}}\left(r\right)}\right)\hspace{-0.1cm}\mathscr{L}_{I_{r}}\hspace{-0.1cm}\left(\frac{\gamma}
{P\zeta_{2}^{\textrm{L}}\left(r\right)}\right)\hspace{-0.1cm}f_{R,2}^{\textrm{L}}\left(r\right)dr.\label{eq:T_2_UAS1_LoS}
\end{equation}

According to Theorem~\ref{thm:p_cov_UAS1} and (\ref{eq:LoS_Prob_func_linear}),
$f_{R,1}^{\textrm{NL}}\left(r\right)$ is given by
\begin{eqnarray}
f_{R,2}^{\textrm{L}}\left(r\right)\hspace{-0.3cm} & = & \hspace{-0.3cm}\exp\left(-\int_{0}^{r_{1}}\lambda\left(1-\textrm{Pr}^{\textrm{L}}\left(u\right)\right)2\pi udu\right)\nonumber \\
\hspace{-0.3cm} &  & \hspace{-0.3cm}\times\exp\left(-\int_{0}^{r}\lambda\textrm{Pr}^{\textrm{L}}\left(u\right)2\pi udu\right)\times0\times2\pi r\lambda\nonumber \\
\hspace{-0.3cm} & = & \hspace{-0.3cm}0,\quad\left(r>d_{1}\right).\label{eq:spec_geom_dis_PDF_UAS1_LoS_seg2}
\end{eqnarray}

Plugging (\ref{eq:spec_geom_dis_PDF_UAS1_LoS_seg2}) into (\ref{eq:T_2_UAS1_LoS}),
yields
\begin{equation}
T_{2}^{\textrm{L}}=0.\label{eq:T_2_UAS1_LoS_final}
\end{equation}

\subsection{The Computation of $T_{2}^{\textrm{NL}}$\label{sub:The-Computation-of-T2NL}}

From Theorem~\ref{thm:p_cov_UAS1},
$T_{2}^{\textrm{NL}}$ can be derived as
\begin{eqnarray}
\hspace{-0.3cm}\hspace{-0.3cm}T_{2}^{\textrm{NL}}\hspace{-0.3cm} & = & \hspace{-0.3cm}\int_{d_{1}}^{\infty}\hspace{-0.1cm}\exp\hspace{-0.1cm}\left(-\frac{\gamma N_{0}}{P\zeta_{2}^{\textrm{NL}}\left(r\right)}\right)\hspace{-0.1cm}\mathscr{L}_{I_{r}}\hspace{-0.1cm}\left(\frac{\gamma}{P\zeta_{2}^{\textrm{NL}}\left(r\right)}\right)\hspace{-0.1cm}f_{R,2}^{\textrm{NL}}\left(r\right)dr\nonumber \\
\hspace{-0.3cm} & \overset{(a)}{=} & \hspace{-0.3cm}\int_{d_{1}}^{\infty}\hspace{-0.1cm}\exp\hspace{-0.1cm}\left(-\frac{\gamma r^{\alpha^{{\rm {NL}}}}N_{0}}{PA^{{\rm {NL}}}}\right)\hspace{-0.1cm}\mathscr{L}_{I_{r}}\hspace{-0.1cm}\left(\frac{\gamma r^{\alpha^{{\rm {NL}}}}}{PA^{{\rm {NL}}}}\right)\hspace{-0.1cm}f_{R,2}^{\textrm{NL}}\left(r\right)dr,\label{eq:T_2_UAS1_NLoS}
\end{eqnarray}
where $\zeta_{2}^{\textrm{NL}}\left(r\right)=A^{{\rm {NL}}}r^{-\alpha^{{\rm {NL}}}}$ from (\ref{eq:PL_BS2UE_2slopes}) is plugged into the step (a) of (\ref{eq:T_2_UAS1_NLoS}).

In (\ref{eq:T_2_UAS1_NLoS}),
according to Theorem~\ref{thm:p_cov_UAS1} and (\ref{eq:LoS_Prob_func_linear}),
%$f_{R,2}^{\textrm{NL}}\left(r\right)$ can be derived as
we have
\begin{eqnarray}
f_{R,2}^{\textrm{NL}}\left(r\right)\hspace{-0.3cm} & = & \hspace{-0.3cm}\exp\left(-\int_{0}^{d_{1}}\lambda\left(1-\frac{u}{d_{1}}\right)2\pi udu\right)\nonumber \\
 &  & \hspace{-0.3cm}\times\exp\left(-\int_{0}^{d_{1}}\lambda\frac{u}{d_{1}}2\pi udu-\int_{d_{1}}^{r}\lambda2\pi udu\right)2\pi r\lambda\nonumber \\
\hspace{-0.3cm} & = & \hspace{-0.3cm}\exp\left(-\pi\lambda r^{2}\right)2\pi r\lambda,\quad\left(r>d_{1}\right).\label{eq:spec_geom_dis_PDF_UAS1_NLoS_seg2}
\end{eqnarray}

Besides, to compute $\mathscr{L}_{I_{r}}\left(\frac{\gamma r^{\alpha^{{\rm {NL}}}}}{PA^{{\rm {NL}}}}\right)$ in (\ref{eq:T_2_UAS1_NLoS}) for the range of $r>d_{1}$,
we propose Lemma~\ref{lem:laplace_term_UAS1_NLoS_seg2}.

\begin{lem}
 \label{lem:laplace_term_UAS1_NLoS_seg2}$\mathscr{L}_{I_{r}}\left(\frac{\gamma r^{\alpha^{{\rm {NL}}}}}{PA^{{\rm {NL}}}}\right)$
in the range of $r>d_{1}$ can be calculated by
\begin{eqnarray}
\mathscr{L}_{I_{r}}\left(\frac{\gamma r^{\alpha^{{\rm {NL}}}}}{PA^{{\rm {NL}}}}\right)\hspace{-0.3cm} & = & \hspace{-0.3cm}\exp\left(-2\pi\lambda\rho_{2}\left(\alpha^{{\rm {NL}}},1,\left(\gamma r^{\alpha^{{\rm {NL}}}}\right)^{-1},r\right)\right),\nonumber \\
\hspace{-0.3cm} &  & \hspace{-0.3cm}\hspace{4cm}\left(r>d_{1}\right),\label{eq:Lemma_5}
\end{eqnarray}
where $\rho_{2}\left(\alpha,\beta,t,d\right)$ is defined in (\ref{eq:rou2_func}).
\end{lem}

\begin{IEEEproof}
See Appendix~D.
\end{IEEEproof}

To sum up, $T_{2}^{\textrm{NL}}$ can be evaluated as
\begin{equation}
\hspace{-0.1cm}\hspace{-0.1cm}T_{2}^{\textrm{NL}}\hspace{-0.1cm}=\hspace{-0.1cm}\int_{d_{1}}^{\infty}\hspace{-0.1cm}\exp\hspace{-0.1cm}\left(-\frac{\gamma r^{\alpha^{{\rm {NL}}}}N_{0}}{PA^{{\rm {NL}}}}\right)\hspace{-0.1cm}\mathscr{L}_{I_{r}}\hspace{-0.1cm}\left(\frac{\gamma r^{\alpha^{{\rm {NL}}}}}{PA^{{\rm {NL}}}}\right)\hspace{-0.1cm}f_{R,2}^{\textrm{NL}}\left(r\right)dr,\label{eq:T_2_UAS1_NLoS_final}
\end{equation}
where $f_{R,2}^{\textrm{NL}}\left(r\right)$ and $\mathscr{L}_{I_{r}}\left(\frac{\gamma r^{\alpha^{{\rm {NL}}}}}{PA^{{\rm {NL}}}}\right)$ are computed by (\ref{eq:spec_geom_dis_PDF_UAS1_NLoS_seg2}) and (\ref{eq:Lemma_5}), respectively.

\subsection{The Results of $p^{\textrm{cov}}\left(\lambda,\gamma\right)$ and $A^{\textrm{ASE}}\left(\lambda,\gamma_{0}\right)$}

Considering (\ref{eq:p_cov_special_case_UAS1}) and bringing together the results from Subsections~\ref{sub:The-Computation-of-T1L}\textasciitilde{}\ref{sub:The-Computation-of-T2NL},
$p^{\textrm{cov}}\left(\lambda,\gamma\right)$ for 3GPP Case~1 can be evaluated as
\begin{equation}
p^{\textrm{cov}}\left(\lambda,\gamma\right)=T_{1}^{\textrm{L}}+T_{1}^{\textrm{NL}}+T_{2}^{\textrm{NL}},\label{eq:spec_p_cov_UAS1_final}
\end{equation}
where $T_{1}^{\textrm{L}}$, $T_{1}^{\textrm{NL}}$ and $T_{2}^{\textrm{NL}}$ are computed from numerically tractable integral-form expressions using (\ref{eq:T_1_UAS1_LoS_final}), (\ref{eq:T_1_UAS1_NLoS_final}) and (\ref{eq:T_2_UAS1_NLoS_final}), respectively.

Plugging $p^{\textrm{cov}}\left(\lambda,\gamma\right)$ obtained from (\ref{eq:spec_p_cov_UAS1_final}) into (\ref{eq:cond_SINR_PDF}),
we can get the result of $A^{\textrm{ASE}}\left(\lambda,\gamma_{0}\right)$
using (\ref{eq:ASE_def}) for 3GPP Case~1.

Regarding the computational process to obtain $p^{\textrm{cov}}\left(\lambda,\gamma\right)$ for 3GPP Case~1,
only one fold of integral is required for the calculation of $\left\{ T_{n}^{\textrm{L}},T_{n}^{\textrm{NL}}\right\} $ in (\ref{eq:spec_p_cov_UAS1_final}),
compared with three folds of integrals for the general case in Theorem~\ref{thm:p_cov_UAS1}.
Note that an additional fold of integral is needed for the calculation of $A^{\textrm{ASE}}\left(\lambda,\gamma_{0}\right)$,
making it a 2-fold integral computation.
Thus, the results for 3GPP Case~1 are much more tractable than those for the general case discussed in Section~\ref{sec:General-Results},
because the \emph{linear} LoS probability function in (\ref{eq:LoS_Prob_func_linear}) permits good tractability.

\section{Simulation and Discussion\label{sec:Simulation-and-Discussion}}

In this section, we use numerical results to establish the accuracy of our analysis and further study the performance of dense SCNs.
According to Tables A.1-3, A.1-4 and A.1-7 of~\cite{TR36.828} and~\cite{SCM_pathloss_model},
we adopt the following parameters for 3GPP Case~1:
$d_{1}=0.3$\ km, $\alpha^{{\rm {L}}}=2.09$, $\alpha^{{\rm {NL}}}=3.75$, $A^{{\rm {L}}}=10^{-10.38}$, $A^{{\rm {NL}}}=10^{-14.54}$, $P=24$\ dBm, $N_{0}=-95$\ dBm
(including a noise figure of 9\ dB at the UE).

\subsection{Validation of the Analytical Results of $p^{\textrm{cov}}\left(\lambda,\gamma\right)$ for 3GPP Case~1\label{sub:Sim-p-cov-3GPP-Case-1}}

For 3GPP Case~1 studied in Section~\ref{sec:A-3GPP-Special-Case},
the results of $p^{\textrm{cov}}\left(\lambda,\gamma\right)$ with $\gamma=0\,\textrm{dB}$ and $\gamma=3\,\textrm{dB}$ are plotted in Fig.~\ref{fig:p_cov_linear_fixedPower24dBm_various_gamma}.
For comparison, we have also included analytical results assuming a single-slope path loss model that does not differentiate LoS and NLoS transmissions~\cite{Jeff's work 2011}.
Note that in~\cite{Jeff's work 2011},
only one path loss exponent is defined and denoted by $\alpha$,
the value of which is assumed to be $\alpha=\alpha^{{\rm {NL}}}=3.75$.
As can be observed from Fig.~\ref{fig:p_cov_linear_fixedPower24dBm_various_gamma},
our analytical results perfectly match the simulation results.
Due to the significant accuracy of $p^{\textrm{cov}}\left(\lambda,\gamma\right)$ and since the results of $A^{\textrm{ASE}}\left(\lambda,\gamma_{0}\right)$ are computed based on $p^{\textrm{cov}}\left(\lambda,\gamma\right)$,
we will only use analytical results of $p^{\textrm{cov}}\left(\lambda,\gamma\right)$ in our discussion hereafter.

\begin{center}
\begin{figure}[H]
\noindent \begin{centering}
\includegraphics[width=8.8cm]{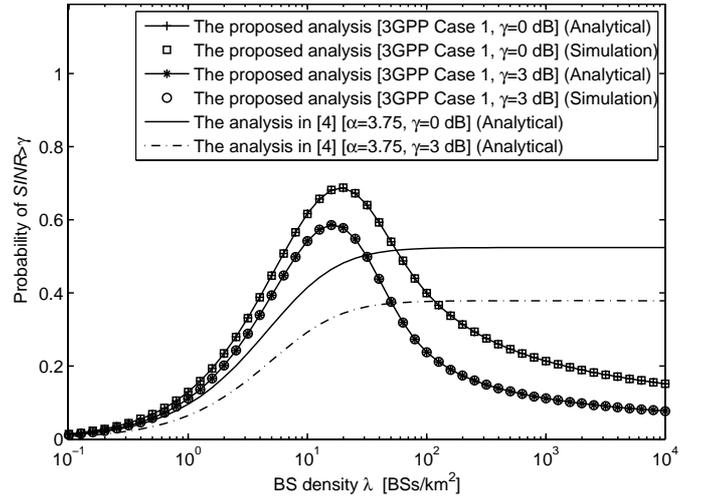}\renewcommand{\figurename}{Fig.}\protect\caption{\label{fig:p_cov_linear_fixedPower24dBm_various_gamma}The coverage
probability $p^{\textrm{cov}}\left(\lambda,\gamma\right)$ vs. the
BS density $\lambda$ for 3GPP Case~1 with various SINR thresholds
$\gamma$.}
\par\end{centering}
\vspace{-0.5cm}
\end{figure}
\par\end{center}
%\vspace{-0.2cm}

From Fig.~\ref{fig:p_cov_linear_fixedPower24dBm_various_gamma},
we can observe that the coverage probability given by~\cite{Jeff's work 2011} first increases with the BS density
because more BSs provide better coverage in noise-limited networks.
Then, when $\lambda$ is large enough, e.g., $\lambda>10^{2}\,\textrm{BSs/km}^{2}$,
the coverage probability becomes independent of $\lambda$ since the network is pushed into the interference-limited region.
The intuition behind the observation is that with the simplistic assumption on the path loss model,
the increase in interference power is counterbalanced by the increase in signal power in a interference-limited network,
and thus the coverage probability remains the same as $\lambda$ further increases~\cite{Jeff's work 2011}.

However, the coverage probability performance of the proposed analysis for 3GPP Case~1 incorporating both LoS and NLoS transmissions exhibits a significant deviation from that of the analysis from~\cite{Jeff's work 2011}.
This is because when the distance $r$ decreases,
or equivalently when the BS density $\lambda$ increases,
LoS transmission occurs with an increasingly higher probability than NLoS transmission.
In more detail,
\begin{itemize}
\item
When the SCN is sparse and thus noise-limited,
e.g., $\lambda\leq10\,\textrm{BSs/km}^{2}$,
the coverage probability given by the proposed analysis grows as $\lambda$ increases.
\item
When the network is dense enough,
e.g., $10\,\textrm{BSs/km}^{2} < \lambda < 10^{3}\,\textrm{BSs/km}^{2}$,
which is the practical range of $\lambda$ for the existing 4G networks and the future 5G network~\cite{Tutor_smallcell},
the coverage probability given by the proposed analysis decreases as $\lambda$ increases,
due to the transition of a large number of interference paths from NLoS to LoS.
Hence, some interfering BSs may be already so close to the typical UE that their signals may start reaching the UE via strong LoS paths.
\item
When the SCN is extremely dense,
e.g., $\lambda\geq10^{3}\,\textrm{BSs/km}^{2}$,
the coverage probability decreases at a  slower pace because both the signal power and the interference power are LoS dominated and thus statistically stable in general.
\end{itemize}

It is important to note that the coverage probability performance of the proposed analysis for 3GPP Case~1 peaks at a certain value $\lambda_{0}$.
Such crucial point can be readily obtained by setting the partial derivative of $p^{\textrm{cov}}\left(\lambda,\gamma\right)$ with regard to $\lambda$ to zero,
i.e., $\lambda^{*}=\underset{\lambda}{\arg}\left\{ \frac{\partial p^{\textrm{cov}}\left(\lambda,\gamma\right)}{\partial\lambda}=0\right\} $.
The solution to this equation can be numerically found using a standard bisection searching~\cite{Bisection}.
In Fig.~\ref{fig:p_cov_linear_fixedPower24dBm_various_gamma},
the numerical results of $\lambda^{*}$ are $19.01\,\textrm{BSs/km}^{2}$ and $16.52\,\textrm{BSs/km}^{2}$ for $\gamma=0\,\textrm{dB}$ and $\gamma=3\,\textrm{dB}$, respectively.

Note that similar trends of the coverage probability vs. the BS density $\lambda$ were also observed in~{[}10-12{]},
i.e., the coverage probability will initially increase with the increase of $\lambda$,
but when $\lambda$ is larger than $\lambda_{0}$, the network coverage probability will decrease as small cells become denser.

Considering such trend of coverage probability performance and looking at the expression of the ASE in (\ref{eq:ASE_def}),
we can conclude that the trend of the ASE performance for SCNs should be complicated,
and it will be investigated in the next subsection.

\subsection{Discussion on the Analytical Results of $A^{\textrm{ASE}}\left(\lambda,\gamma_{0}\right)$ for 3GPP Case~1\label{sub:Sim-ASE-3GPP-Case-1}}

In this subsection, we investigate the analytical results of $A^{\textrm{ASE}}\left(\lambda,\gamma_{0}\right)$ with $\gamma_{0}=0,3,6\,\textrm{dB}$ based on the analytical results of $p^{\textrm{cov}}\left(\lambda,\gamma\right)$.
Our results of $A^{\textrm{ASE}}\left(\lambda,\gamma_{0}\right)$ are plotted in Fig.~\ref{fig:ASE_linear_fixedPower24dBm_various_gamma},
comparing with the analytical results from~\cite{Jeff's work 2011} with $\gamma_{0}=0\,\textrm{dB}$.
\vspace{-0.2cm}
\begin{center}
\begin{figure}[H]
\noindent \begin{centering}
\includegraphics[width=8.8cm]{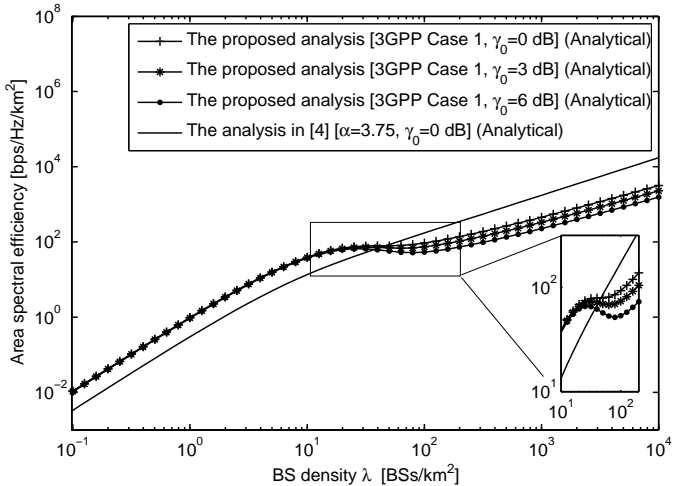}\renewcommand{\figurename}{Fig.}\protect\caption{\label{fig:ASE_linear_fixedPower24dBm_various_gamma}The ASE $A^{\textrm{ASE}}\left(\lambda,\gamma_{0}\right)$
vs. the BS density $\lambda$ for 3GPP Case~1 with various SINR thresholds
$\gamma_{0}$.}
\par\end{centering}
\vspace{-0.5cm}
\end{figure}
\par\end{center}
%\vspace{-0.2cm}

As can be seen from Fig.~\ref{fig:ASE_linear_fixedPower24dBm_various_gamma},
the analysis from~\cite{Jeff's work 2011} indicates that when the SCN is dense enough,
e.g., $\lambda\geq10^{2}\,\textrm{BSs/km}^{2}$,
the ASE performance increases linearly with $\lambda$,
which is logically correct from the conclusion that $p^{\textrm{cov}}\left(\lambda,\gamma\right)$ is invariable with respect to $\lambda$ for a given $\gamma$
when $\lambda$ is sufficiently large~\cite{Jeff's work 2011}.

In contrast, our proposed analysis for 3GPP Case~1 reveals a more complicated trend for the ASE performance,
which is highlighted by the sub-figure in Fig.~\ref{fig:ASE_linear_fixedPower24dBm_various_gamma}.
In more detail,
\begin{itemize}
\item
When the SCN is sparse and thus noise-limited,
e.g., $\lambda\leq\lambda_{0}\,\textrm{BSs/km}^{2}$ ($\lambda_{0}\approx20\,\textrm{BSs/km}^{2}$ in Fig.~\ref{fig:ASE_linear_fixedPower24dBm_various_gamma}),
the ASE quickly increases with $\lambda$ because the network is generally noise-limited,
and thus adding more small cells immensely benefits the ASE.
\item
When the network is dense enough,
i.e., $\lambda\in\left[\lambda_{0},10^{3}\right]\,\textrm{BSs/km}^{2}$,
which is the practical range of $\lambda$ for the existing 4G networks and the future 5G network~\cite{Tutor_smallcell},
%(practical range of $\lambda$ for the existing 4G networks and the future 5G networks),
the trend of the ASE performance is very interesting.
First, when $\lambda\in\left[\lambda_{0},\lambda_{1}\right]\,\textrm{BSs/km}^{2}$,
where $\lambda_{1}$ is another threshold larger than $\lambda_{0}$ ($\lambda_{1}\approx10^{2}\,\textrm{BSs/km}^{2}$ in Fig.~\ref{fig:ASE_linear_fixedPower24dBm_various_gamma}),
the ASE exhibits a slowing-down in the rate of growth ($\gamma_{0}=0\,\textrm{dB}$) or even a decrease ($\gamma_{0}=3,6\,\textrm{dB}$)
due to the fast decrease of the coverage probability at $\lambda\in\left[\lambda_{0},\lambda_{1}\right]\,\textrm{BSs/km}^{2}$,
as shown in Fig.~\ref{fig:p_cov_linear_fixedPower24dBm_various_gamma}.
Second, when $\lambda\geq\lambda_{1}$,
the ASE will pick up the growth rate since the decrease of the coverage probability becomes a minor factor compared with the increase of $\lambda$.
\item
When the SCN is extremely dense,
e.g., $\lambda$ is larger than $10^{3}\,\textrm{BSs/km}^{2}$,
the ASE exhibits a nearly linear trajectory with regard to $\lambda$ because both the signal power and the interference power are now LoS dominated,
and thus statistically stable as explained before.
\end{itemize}

In particular, our finding that the ASE may decrease as the BS density increases for the practical range of $\lambda$ for the existing 4G networks and the future 5G networks~\cite{Tutor_smallcell}
indicates the significant impact of the path loss model incorporating both NLoS and LoS transmissions on the ASE performance.
Such impact makes a difference for the ASE of dense SCNs both quantitatively and qualitatively compared with analyses with simplistic path loss models that does not differentiate LoS and NLoS transmissions.
As a confirmation, note that in Fig.~\ref{fig:p_cov_linear_fixedPower24dBm_various_gamma},
we can observe that increasing the SINR threshold $\gamma$ from 0\,dB to 3\,dB will accelerate the decrease of the coverage probability at $\lambda\in\left[\lambda_{0},\lambda_{1}\right]\,\textrm{BSs/km}^{2}$ because of the more demanding SINR requirement,
which in turn causes the decrease of the ASE ($\gamma_{0}=3,6\,\textrm{dB}$) at that range of $\lambda$ in Fig.~\ref{fig:ASE_linear_fixedPower24dBm_various_gamma}.
Note that our conclusion is made from the investigated set of parameters,
and it is of significant interest to further study the generality of this conclusion in other network models and with other parameter sets.

To sum up, our results are different from those in existing studies~{[}10-12{]} and those in~\cite{Jeff's work 2011} assuming a simplistic path loss model that does not differentiate LoS and NLoS transmissions.
The implication is profound,
which is summarized in the following remarks:
\begin{itemize}
\item
\textbf{Remark}~\textbf{1: }
From the investigated set of parameters,
we find that when the density of small cells is larger than a threshold $\lambda_{0}$,
the ASE may suffer from a slow growth or even a decrease as the BS density increases because of the quick decrease of the network coverage probability.
It is of significant interest to further study the generality of this conclusion in other network models and with other parameter sets.
\item
\textbf{Remark}~\textbf{2: }
Our finding is of significant importance for practical SCN deployments in the following two aspects:
\emph{(i)} The valley area,
where the BS density $\lambda$ is tens to hundreds of $\textrm{BSs/km}^{2}$ and the ASE may suffer from a slow growth or even a decrease,
stands in our way of the \emph{evolution} from 4G to 5G.
Since $\lambda$ has been estimated as several to tens of $\textrm{BSs/km}^{2}$ in 4G~\cite{TR36.828},~\cite{Book_LTE2}
and tens to hundreds or even thousands of $\textrm{BSs/km}^{2}$ in 5G~\cite{Tutor_smallcell},
how to cost-efficiently march cross this undesirable valley area using new technologies is crucial for the commercial success of future 4G/5G networks.
\emph{(ii)} Our results are not obtained on conditions such as near-field path loss exponents,
i.e., $\alpha^{{\rm {L}}}<2$ or $\alpha^{{\rm {NL}}}<2$.
All the parameters are practical ones recommended by the 3GPP standards for the state-of-the-art SCNs.
Therefore, our conclusion can provide some valuable guidance for operators in their quest of network densification.
%when they want to densify their telecommunication networks.
\item
\textbf{Remark}~\textbf{3: }
The ASE will grow almost linearly as the BS density increases above $\lambda_{1}$,
which might be a candidate BS density threshold for characterizing the ultra-dense SCNs in future 5G networks~\cite{Tutor_smallcell}.
\end{itemize}

\subsection{Discussion on Various Values of $\alpha^{{\rm {L}}}$ for 3GPP Case~1\label{sub:alpha-LoS-3GPP-Case-1}}

As discussed in Subsection~\ref{sub:Sim-ASE-3GPP-Case-1},
it is of interest to further study the generality of our conclusion in other network models and with other parameter sets.
In this subsection,
we change the value of $\alpha^{{\rm {L}}}$ from 2.09 to 1.09 and 3.09 to investigate the impact of $\alpha^{{\rm {L}}}$ on our conclusion.
In Fig.~\ref{fig:ASE_linear_fixedPower24dBm_various_alphaLoS},
the analytical results of $A^{\textrm{ASE}}\left(\lambda,\gamma_{0}\right)$ with $\gamma_{0}=0\,\textrm{dB}$ and various values of $\alpha^{{\rm {L}}}$ are compared
with the results from~\cite{Jeff's work 2011} ($\alpha=3.75$).
%Note that since the analysis in~\cite{Jeff's work 2011} does not differentiate LoS and NLoS transmissions, the results from~\cite{Jeff's work 2011} are equivalent to our results with $\alpha^{{\rm {L}}}=\alpha^{{\rm {NL}}}=3.75$.
%[David]: I think should be NL here. I have changed $\alpha^{{\rm {L}}}=3.75$ to  $\alpha^{{\rm {NL}}}=3.75$
\vspace{-0.2cm}
\noindent \begin{center}
\begin{figure}[H]
\noindent \begin{centering}
\includegraphics[width=8.8cm]{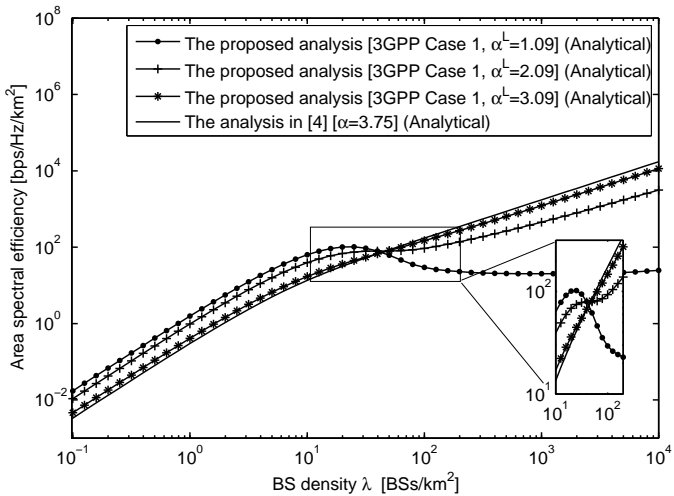}\renewcommand{\figurename}{Fig.}\protect\caption{\label{fig:ASE_linear_fixedPower24dBm_various_alphaLoS}The ASE $A^{\textrm{ASE}}\left(\lambda,\gamma_{0}\right)$
vs. the BS density $\lambda$ for 3GPP Case~1 with SINR threshold
$\gamma_{0}=0\,\textrm{dB}$ and with various LoS path loss exponents
$\alpha^{{\rm {L}}}$. }
\par\end{centering}
\vspace{-0.5cm}
\end{figure}
\par\end{center}

As can be seen from Fig.~\ref{fig:ASE_linear_fixedPower24dBm_various_alphaLoS},
the slow growth or the decrease of ASE at $\lambda\in\left[\lambda_{0},\lambda_{1}\right]\,\textrm{BSs/km}^{2}$
($\lambda_{0}\approx20\,\textrm{BSs/km}^{2}$ and $\lambda_{1}\approx10^{2}\,\textrm{BSs/km}^{2}$)
is more obvious when the difference between the NLoS and the LoS path loss exponents, $\alpha^{{\rm {NL}}}$ and $\alpha^{{\rm {L}}}$, is larger.
This is because the transition of interference from NLoS transmission to LoS transmission becomes more drastic.
For example, when $\alpha^{{\rm {L}}}$ takes a near-field path loss exponent such as 1.09~\cite{related_work_Jeff},
the decrease of ASE at $\lambda\in\left[\lambda_{0},\lambda_{1}\right]\,\textrm{BSs/km}^{2}$ is significant,
and it hardly recovers after $\lambda_{1}$.
Such observation is in agreement with that in~\cite{related_work_Jeff} when $\alpha^{{\rm {L}}}<2$.
However, the difference between our results and those in~\cite{related_work_Jeff} is that
we have shown how the ASE may also decrease even when $\alpha^{{\rm {L}}}\geq2$.
Such difference is due to the different definition of ASE in~\cite{related_work_Jeff} and in our paper.
Note that in~\cite{related_work_Jeff},
a deterministic rate based on $\gamma_{0}$ is assumed for the typical UE,
no matter what the actual SINR value is.
Our definition of ASE in (\ref{eq:ASE_def}) is more realistic due to the SINR-dependent rate,
but it is more difficult to analyze,
as it requires one more fold of numerical integral compared with~\cite{related_work_Jeff}.
To sum up, we draw the following remark from the discussion of this subsection.
\begin{itemize}
\item
\textbf{Remark}~\textbf{4: }
We observe that the behavior of the ASE depends on the characteristics of the LoS and the NLoS path loss functions.
The larger the difference between the LoS and the NLoS path loss exponents,
the more the ASE suffers at $\lambda\in\left[\lambda_{0},\lambda_{1}\right]\,\textrm{BSs/km}^{2}$ due to more drastic transition of interference from the NLoS transmission to the LoS transmission.
\end{itemize}

\subsection{Investigation of 3GPP Case~2\label{sub:Investigation-3GPP-Case-2}}

In this subsection, we investigate the ASE performance for 3GPP Case~2,
which has been discussed in Section~\ref{sec:A-3GPP-Special-Case}.
Note that the parameters of the LoS probability function $\textrm{Pr}^{\textrm{L}}\left(r\right)$ for 3GPP Case~2 are $R_{1}=0.156$\ km and $R_{2}=0.03$\ km~\cite{TR36.828}.
Due to the complicated expressions of (\ref{eq:LoS_Prob_func_reverseS_shape}),
numerically tractable integral-form expressions for $p^{\textrm{cov}}\left(\lambda,\gamma\right)$ like those in (\ref{eq:p_cov_special_case_UAS1}) are difficult to obtain.
Instead, we directly apply numerical integration in Theorem~\ref{thm:p_cov_UAS1} to evaluate the analytical results for 3GPP Case~2.
It is important to note that as discussed in Section~\ref{sec:General-Results} and Section~\ref{sec:A-3GPP-Special-Case},
3GPP Case~1 only requires one and two folds of integrals to obtain $p^{\textrm{cov}}\left(\lambda,\gamma\right)$ and $A^{\textrm{ASE}}\left(\lambda,\gamma_{0}\right)$,
while 3GPP Case~2 requires three and four folds of integrals to compute $p^{\textrm{cov}}\left(\lambda,\gamma\right)$ and $A^{\textrm{ASE}}\left(\lambda,\gamma_{0}\right)$.

Besides, in order to show the versatility of the studied 3GPP Case~1 with the \emph{linear} LoS probability function shown in (\ref{eq:LoS_Prob_func_linear}),
we propose to use a piece-wise \emph{linear} LoS probability function to approximate the complicated LoS probability function of 3GPP Case~2
given by (\ref{eq:LoS_Prob_func_reverseS_shape}). Our proposed approximation of the LoS probability function for 3GPP Case~2 is defined as a 3-piece linear function as
%\begin{singlespace}
%\noindent
\begin{equation}
\textrm{Pr}^{\textrm{L}}\left(r\right)=\begin{cases}
\begin{array}{l}
1,\\
1-\frac{r-d_{1}}{d_{2}-d_{1}},\\
0,
\end{array} & \begin{array}{l}
0<r\leq d_{1}\\
d_{1}<r\leq d_{2}\\
r>d_{2}
\end{array}\end{cases},\label{eq:LoS_Prob_func_approx_reverseS}
\end{equation}
%\end{singlespace}
where $d_{1}$ and $d_{2}$ are set to 0.0184\ km and 0.1171\ km, respectively.
Note that $d_{1}$ is chosen as 0.0184\ km because $\textrm{Pr}^{\textrm{L}}\left(0.0184\right)=0.999\approx1$ in (\ref{eq:LoS_Prob_func_reverseS_shape}).
Besides, the value of $d_{2}$ is chosen as 0.1171\ km because
$\textrm{Pr}^{\textrm{L}}\left(r\right)$ in (\ref{eq:LoS_Prob_func_approx_reverseS}) needs to go through point $\left(\frac{0.156}{\ln10},0.5\right)$,
which is the point connecting the two segments in $\textrm{Pr}^{\textrm{L}}\left(r\right)$ of 3GPP Case~2 given by (\ref{eq:LoS_Prob_func_reverseS_shape}).

Fig.~\ref{fig:Pr_LoS_3GPP_Case-2} illustrates $\textrm{Pr}^{\textrm{L}}\left(r\right)$ as defined by 3GPP Case~2 in (\ref{eq:LoS_Prob_func_reverseS_shape})
and as approximated by a piece-wise linear function shown in (\ref{eq:LoS_Prob_func_approx_reverseS}).
For clarity, the combined case with the path loss function in (\ref{eq:PL_BS2UE_2slopes}) and the 3-piece LoS probability function in (\ref{eq:LoS_Prob_func_approx_reverseS})
is referred to as the Approximated 3GPP Case~2.
Note that the accuracy of our approximation can be easily improved by fitting the LoS probability function with more than 3 pieces.
Based on Theorem~\ref{thm:p_cov_UAS1},
we can readily extend the results in Section~\ref{sec:A-3GPP-Special-Case} to analyze the Approximated 3GPP Case~2 in a tractable manner.
The details are omitted for brevity.
\vspace{-0.2cm}
\begin{center}
\begin{figure}[H]
\noindent \begin{centering}
\includegraphics[width=8.8cm]{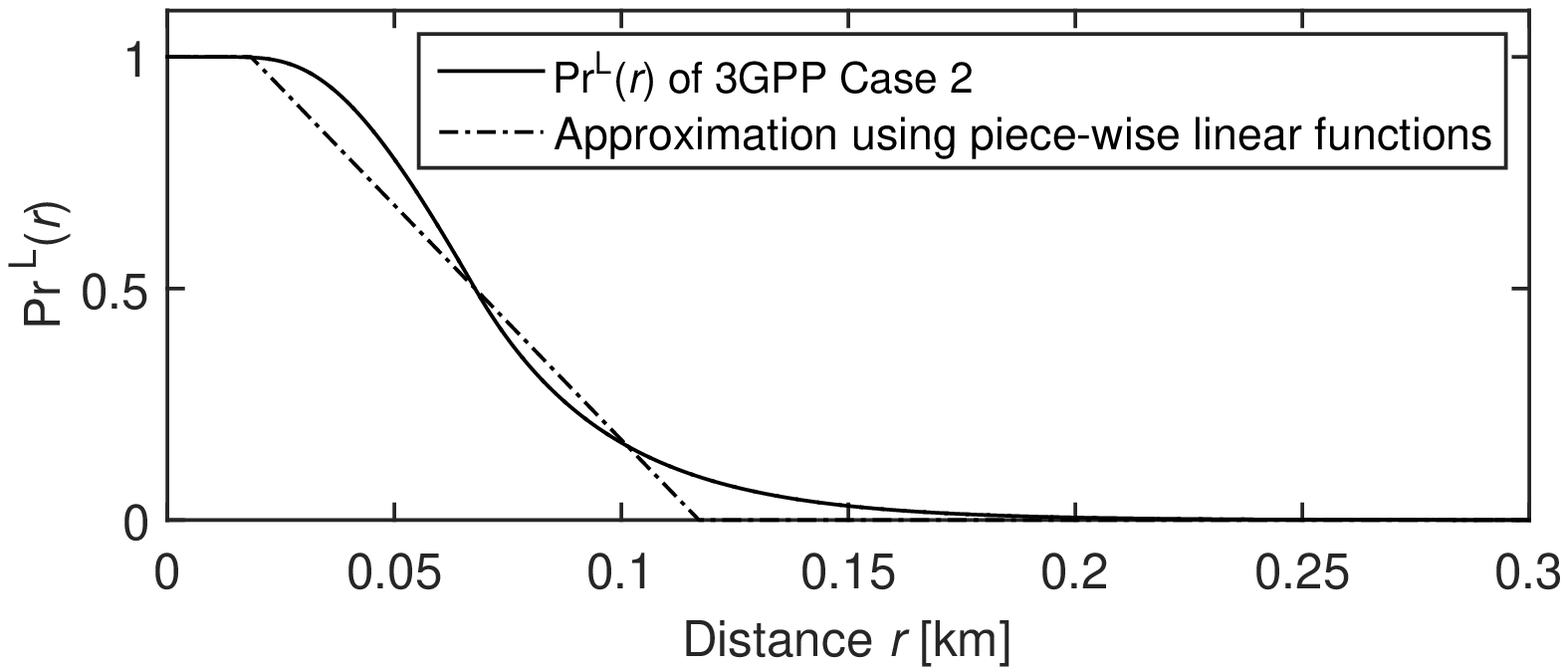}\renewcommand{\figurename}{Fig.}\protect\caption{\label{fig:Pr_LoS_3GPP_Case-2}The LoS probability $\textrm{Pr}^{\textrm{L}}\left(r\right)$ vs. the distance $r$ for 3GPP Case 2~\cite{TR36.828}.}
\par\end{centering}
\vspace{-0.5cm}
\end{figure}
\par\end{center}

The results of $A^{\textrm{ASE}}\left(\lambda,\gamma_{0}\right)$ are plotted in Fig.~\ref{fig:ASE_reverseS_fixedPower24dBm_various_gamma}.
As can be seen from it,
the results of the Approximated 3GPP Case~2 match those of 3GPP Case~2 well,
thus showing the accuracy and the usefulness of our analysis with the \emph{linear} LoS probability function in Section~\ref{sec:A-3GPP-Special-Case}.
More importantly, all the observations in Subsection~\ref{sub:Sim-ASE-3GPP-Case-1}
are qualitatively valid for Fig.~\ref{fig:ASE_reverseS_fixedPower24dBm_various_gamma}
except for some quantitative deviation.

\vspace{-0.2cm}
\noindent \begin{center}
\begin{figure}[H]
\noindent \begin{centering}
\includegraphics[width=8.8cm]{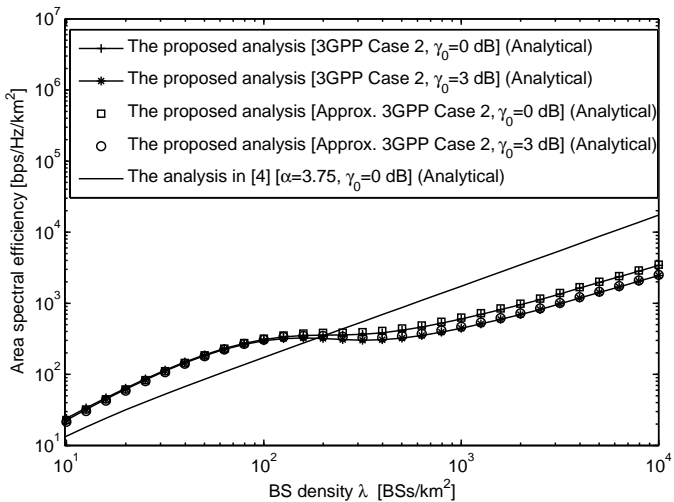}\renewcommand{\figurename}{Fig.}\protect\caption{\label{fig:ASE_reverseS_fixedPower24dBm_various_gamma}The ASE $A^{\textrm{ASE}}\left(\lambda,\gamma_{0}\right)$
vs. the BS density $\lambda$ for 3GPP Case~2 with various SINR thresholds
$\gamma_{0}$.}
\par\end{centering}
\vspace{-0.5cm}
\end{figure}
\par\end{center}

Note that 3GPP Case~2 was treated in~\cite{related_work_Galiotto} by approximating $\textrm{Pr}^{\textrm{L}}\left(r\right)$ in (\ref{eq:LoS_Prob_func_reverseS_shape})
as an exponentially decreasing function.
However, the results in~\cite{related_work_Galiotto} are less tractable than our approximation.
The former results require three and four folds of integrals to obtain $p^{\textrm{cov}}\left(\lambda,\gamma\right)$ and $A^{\textrm{ASE}}\left(\lambda,\gamma_{0}\right)$,
while our approximation only require one and two folds of integrals to obtain $p^{\textrm{cov}}\left(\lambda,\gamma\right)$ and $A^{\textrm{ASE}}\left(\lambda,\gamma_{0}\right)$, thanks to the good tractability provided by the \emph{linear} LoS probability function.

%[David]: I don't think we need this since it is repetitive.
%Specifically, in Fig.~\ref{fig:ASE_reverseS_fixedPower24dBm_various_gamma},
%the numerical result for $\lambda_{0}$ is around $10^{2}\,\textrm{BSs/km}^{2}$.
%In Fig.~\ref{fig:ASE_reverseS_fixedPower24dBm_various_gamma},
%the ASE is also suffers from a slow growth ($\gamma_{0}=0\,\textrm{dB}$) or a decrease ($\gamma_{0}=3\,\textrm{dB}$) as $\lambda$ increases when $\lambda>\lambda_{0}$
%due to the transition of a large number of interference paths from NLoS to LoS as explained earlier.
%Furthermore, the ASE will grow almost linearly as $\lambda$ increases above another larger threshold $\lambda_{1}$.
%Such $\lambda_{1}$ is in the order of several hundreds of $\textrm{BSs/km}^{2}$ as shown in Fig.~\ref{fig:ASE_reverseS_fixedPower24dBm_various_gamma}.

%Therefore, from the investigated set of parameters for 3GPP Case~2,
%the trend observed on the ASE performance remains the same as that of 3GPP Case~1.
In the following, we summarize our study in this subsection.
\begin{itemize}
\item \noindent \textbf{Remark}~\textbf{5: }
The approach of approximating a LoS probability function as a piece-wise \emph{linear} function
and invoking the results derived in Section~\ref{sec:A-3GPP-Special-Case},
can deal with complicated path loss models in a tractable manner,
which shows the usefulness of studying 3GPP Case~1 with the \emph{linear} LoS probability function.
\end{itemize}

\section{Conclusion\label{sec:Conclusion}}

In this paper, we show that a path loss model incorporating both LoS and NLoS transmissions has a significant impact on the ASE performance of SCNs.
Such impact is not only quantitative but also qualitative.
Previous results predicted that the ASE should monotonically grow with the increase of the BS density.
Instead, our theoretical analysis concludes that when the BS density is larger than a threshold,
the ASE may suffer from a slow growth or even a decrease
as the BS density further increases.
The intuition behind our conclusion is that when the density of small cells is larger than a threshold,
the interference power will increase faster than the signal power due to the transition of a large number of interference paths from NLoS to LoS.
Therefore, our results show that the BS density matters regarding the ASE performance,
thus shedding valuable insights on the deployments of future dense SCNs.

As our future work, we will consider other factors of realistic networks in the theoretical analysis for SCNs,
such as practical directional antennas.
Another future work is the introduction of Rician fading or Nakagami fading in our analysis
because the multi-path fading model is also affected by the LoS and NLoS transmissions.

\section*{Acknowlegements}

The authors would like to thank all the anonymous reviewers for their
helpful comments and constructive suggestions to improve early drafts
of this paper.

\section*{Appendix~A: Proof of Theorem~\ref{thm:p_cov_UAS1} \label{sec:Appendix-A}}

For clarity, we first summarize our ideas to prove Theorem~\ref{thm:p_cov_UAS1}.
In order to evaluate $p^{\textrm{cov}}\left(\lambda,\gamma\right)$,
the first key step is to calculate the distance PDFs for the events
that the typical UE is associated with a BS with a LoS path or that
with a NLoS path, so that the integral of $\textrm{Pr}\left[\mathrm{SINR}>\gamma\right]$
can be performed over the distance $r$. The second key step is to
calculate $\textrm{Pr}\left[\mathrm{SINR}>\gamma\right]$ for the
LoS and the NLoS cases conditioned on $r$.

From (\ref{eq:Coverage_Prob_def}) and (\ref{eq:SINR}), we can derive
$p^{\textrm{cov}}\left(\lambda,\gamma\right)$ as
\begin{eqnarray}
\hspace{-0.3cm}\hspace{-0.3cm}p^{\textrm{cov}}\left(\lambda,\gamma\right)\hspace{-0.3cm} & \stackrel{\left(a\right)}{=} & \hspace{-0.3cm}\int_{r>0}\textrm{Pr}\left[\left.\mathrm{SINR}>\gamma\right|r\right]f_{R}\left(r\right)dr\nonumber \\
\hspace{-0.3cm} & = & \hspace{-0.3cm}\int_{r>0}\textrm{Pr}\left[\frac{P\zeta\left(r\right)h}{I_{r}+N_{0}}>\gamma\right]f_{R}\left(r\right)dr\nonumber \\
\hspace{-0.3cm} & = & \hspace{-0.3cm}\int_{0}^{d_{1}}\textrm{Pr}\left[\frac{P\zeta_{1}^{\textrm{L}}\left(r\right)h}{I_{r}+N_{0}}>\gamma\right]\hspace{-0.1cm}f_{R,1}^{\textrm{L}}\left(r\right)dr\nonumber \\
 &  & \hspace{-0.3cm}+\int_{0}^{d_{1}}\textrm{Pr}\left[\frac{P\zeta_{1}^{\textrm{NL}}\left(r\right)h}{I_{r}+N_{0}}>\gamma\right]\hspace{-0.1cm}f_{R,1}^{\textrm{NL}}\left(r\right)dr\nonumber \\
\hspace{-0.3cm} &  & \hspace{-0.3cm}+\cdots\nonumber \\
\hspace{-0.3cm} &  & \hspace{-0.3cm}+\int_{d_{N-1}}^{\infty}\textrm{Pr}\left[\frac{P\zeta_{N}^{\textrm{L}}\left(r\right)h}{I_{r}+N_{0}}>\gamma\right]\hspace{-0.1cm}f_{R,N}^{\textrm{L}}\left(r\right)dr\nonumber \\
 &  & \hspace{-0.3cm}+\int_{d_{N-1}}^{\infty}\textrm{Pr}\left[\frac{P\zeta_{N}^{\textrm{NL}}\left(r\right)h}{I_{r}+N_{0}}>\gamma\right]\hspace{-0.1cm}f_{R,N}^{\textrm{NL}}\left(r\right)dr\nonumber \\
\hspace{-0.3cm} & \stackrel{\bigtriangleup}{=} & \hspace{-0.3cm}\sum_{n=1}^{N}\left(T_{n}^{\textrm{L}}+T_{n}^{\textrm{NL}}\right),\label{eq:p_cov_general_form}
\end{eqnarray}
where $f_{R,n}^{\textrm{L}}\left(r\right)$
and $f_{R,n}^{\textrm{NL}}\left(r\right)$ are the piece-wise PDF
of the RVs $R_{n}^{\textrm{L}}$ and $R_{n}^{\textrm{NL}}$, where
$R_{n}^{\textrm{L}}$ and $R_{n}^{\textrm{NL}}$ are the distance
that the UE is connected to a BS with a LoS path and the distance
that the UE is connected to a BS with a NLoS path, respectively.
Note that the two events that the typical UE can be successfully covered by a
BS with a LoS path and that with a NLoS path are disjoint events.
Hence, the coverage probability is the direct sum of these two probabilities,
which should be computed based on
$f_{R,n}^{\textrm{L}}\left(r\right)$
and $f_{R,n}^{\textrm{NL}}\left(r\right)$, respectively.
%correspond to those two disjoint events, respectively.
Moreover, $T_{n}^{\textrm{L}}$
and $T_{n}^{\textrm{NL}}$ are piece-wise functions defined as $T_{n}^{\textrm{L}}=\int_{d_{n-1}}^{d_{n}}\textrm{Pr}\left[\frac{P\zeta_{n}^{\textrm{L}}\left(r\right)h}{I_{r}+N_{0}}>\gamma\right]f_{R,n}^{\textrm{L}}\left(r\right)dr$
and $T_{n}^{\textrm{NL}}=\int_{d_{n-1}}^{d_{n}}\textrm{Pr}\left[\frac{P\zeta_{n}^{\textrm{NL}}\left(r\right)h}{I_{r}+N_{0}}>\gamma\right]f_{R,n}^{\textrm{NL}}\left(r\right)dr$,
respectively. Besides, $d_{0}$ and $d_{N}$ are respectively defined
as $0$ and $\infty$.

All the $f_{R,n}^{\textrm{L}}\left(r\right)$ and $f_{R,n}^{\textrm{NL}}\left(r\right)$
are stacked into $f_{R}\left(r\right)$ shown in the step (a) of (\ref{eq:p_cov_general_form}),
and $f_{R}\left(r\right)$ is formally defined in (\ref{eq:general_geom_density}) shown on the top of next page,
which takes a similar form as~(\ref{eq:prop_PL_model}). Since the
two events that the typical UE is connected to a BS with a LoS
path and that with a NLoS path are disjoint events, we have
$f_{R,n}\left(r\right)=f_{R,n}^{\textrm{L}}\left(r\right)+f_{R,n}^{\textrm{NL}}\left(r\right),\forall n\in\left\{ 1,\ldots,n\right\}$
and $\sum_{n=1}^{N}\int_{d_{n-1}}^{d_{n}}f_{R,n}\left(r\right)dr=1$.

%\noindent
\begin{algorithm*}
\begin{equation}
f_{R}\left(r\right)=\begin{cases}
f_{R,1}\left(r\right)=\begin{cases}
\begin{array}{l}
f_{R,1}^{\textrm{L}}\hspace{-0.1cm}\left(r\right),\\
f_{R,1}^{\textrm{NL}}\hspace{-0.1cm}\left(r\right),
\end{array} & \begin{array}{l}
\textrm{when the UE is associated with a LoS BS}\\
\textrm{when the UE is associated with a NLoS BS}
\end{array}\end{cases}, & 0\leq r\leq d_{1}\\
f_{R,2}\left(r\right)=\begin{cases}
\begin{array}{l}
f_{R,2}^{\textrm{L}}\hspace{-0.1cm}\left(r\right),\\
f_{R,2}^{\textrm{NL}}\hspace{-0.1cm}\left(r\right),
\end{array} & \begin{array}{l}
\textrm{when the UE is associated with a LoS BS}\\
\textrm{when the UE is associated with a NLoS BS}
\end{array}\end{cases}, & d_{1}<r\leq d_{2}\\
\vdots & \vdots\\
f_{R,N}\left(r\right)=\begin{cases}
\begin{array}{l}
f_{R,N}^{\textrm{L}}\hspace{-0.1cm}\left(r\right),\\
f_{R,N}^{\textrm{NL}}\hspace{-0.1cm}\left(r\right),
\end{array} & \begin{array}{l}
\textrm{when the UE is associated with a LoS BS}\\
\textrm{when the UE is associated with a NLoS BS}
\end{array}\end{cases}, & r>d_{N-1}
\end{cases}.\label{eq:general_geom_density}
\end{equation}
\end{algorithm*}

In the following, we show how to compute $f_{R,n}^{\textrm{L}}\left(r\right)$
in~(\ref{eq:p_cov_general_form}). To that end, we define two events
as follows.
\begin{itemize}
\item \textbf{Event $B^{\textrm{L}}$: The nearest BS with a LoS path to the typical UE,
is located at distance $X^{\textrm{L}}$.} According to~\cite{Jeff's work 2011},
the CCDF of $X^{\textrm{L}}$ is written as $\bar{F}_{X}^{\textrm{L}}\left(x\right)=\exp\left(-\int_{0}^{x}\textrm{Pr}^{\textrm{L}}\left(u\right)2\pi u\lambda du\right)$.
Taking the derivative of $\left(1-\bar{F}_{X}^{\textrm{L}}\left(x\right)\right)$
with regard to $x$, we can get the PDF of $X^{\textrm{L}}$ as\\
\begin{equation}
\hspace{-0.3cm}f_{X}^{\textrm{L}}\left(x\right)=\exp\left(-\int_{0}^{x}\textrm{Pr}^{\textrm{L}}\left(u\right)2\pi u\lambda du\right)\textrm{Pr}^{\textrm{L}}\left(x\right)2\pi x\lambda.\label{eq:PDF_X_BL}
\end{equation}

\item \textbf{Event $C^{\textrm{NL}}$ conditioned on the value of $X^{\textrm{L}}$:
Given that $X^{\textrm{L}}=x$, the typical UE is then associated with such BS at distance $X^{\textrm{L}}=x$.}
To make the typical UE associated with the LoS BS at distance $X^{\textrm{L}}=x$,
such BS should give the smallest path loss (i.e., the largest $\zeta\left(r\right)$)
from such BS to the typical UE, i.e.,
there should be no BS
with a NLoS path inside the disk centered on the UE with a radius
of $x_{1}<x$ to outperform such LoS BS at distance $X^{\textrm{L}}=x$,
where $x_{1}$ satisfies $x_{1}=\underset{x_{1}}{\arg}\left\{ \zeta^{\textrm{NL}}\left(x_{1}\right)=\zeta^{\textrm{L}}\left(x\right)\right\} $.
According to~\cite{Jeff's work 2011}, such conditional probability
of $C^{\textrm{NL}}$ on condition of $X^{\textrm{L}}=x$ can be computed by
\begin{equation}
\hspace{-0.1cm}\hspace{-0.1cm}\textrm{Pr}\left[\left.C^{\textrm{NL}}\right|X^{\textrm{L}}=x\right]=\exp\hspace{-0.1cm}\left(\hspace{-0.1cm}-\hspace{-0.1cm}\int_{0}^{x_{1}}\hspace{-0.1cm}\hspace{-0.1cm}\left(1-\textrm{Pr}^{\textrm{L}}\left(u\right)\right)2\pi u\lambda du\right)\hspace{-0.1cm}.\label{eq:cond_prob_CNL}
\end{equation}

\end{itemize}

Note that Event $B^{\textrm{L}}$ guarantees that the path loss value
$\zeta^{\textrm{L}}\left(r\right)$ associated with \emph{an arbitrary
LoS BS} is always smaller than that associated with \emph{the considered
LoS BS} at distance $x$. Besides, conditioned on $X^{\textrm{L}}=x$,
Event $C^{N\textrm{L}}$ guarantees that the path loss value $\zeta^{\textrm{NL}}\left(r\right)$
associated with \emph{an arbitrary NLoS BS} is always smaller than
that associated with \emph{the considered LoS BS} at distance $x$.

Then, we consider an unconditional event that the typical UE is associated with
a BS with a LoS path and such BS is located at distance $R^{\textrm{L}}$.
The CCDF of $R^{\textrm{L}}$, denoted by $\bar{F}_{R}^{\textrm{L}}\left(r\right)$,
can be derived as
\begin{eqnarray}
\hspace{-0.8cm}\bar{F}_{R}^{\textrm{L}}\left(r\right)\hspace{-0.3cm} & = & \hspace{-0.3cm}\textrm{Pr}\left[R^{\textrm{L}}>r\right]\nonumber \\
\hspace{-0.3cm} & \stackrel{\left(a\right)}{=} & \hspace{-0.3cm}\textrm{E}_{\left[X^{\textrm{L}}\right]}\left\{ \textrm{Pr}\left[\left.R^{\textrm{L}}>r\right|X^{\textrm{L}}\right]\right\}\hspace{3cm} \nonumber
\end{eqnarray}
\begin{eqnarray}
\hspace{-0.3cm} & = & \hspace{-0.3cm}\int_{0}^{+\infty}\textrm{Pr}\left[\left.R^{\textrm{L}}>r\right|X^{\textrm{L}}=x\right]f_{X}^{\textrm{L}}\left(x\right)dx\nonumber \\
\hspace{-0.3cm} & \stackrel{\left(b\right)}{=} & \hspace{-0.3cm}\int_{0}^{r}\hspace{-0.1cm}0\times f_{X}^{\textrm{L}}\left(x\right)dx+\hspace{-0.1cm}\int_{r}^{+\infty}\hspace{-0.1cm}\hspace{-0.1cm}\hspace{-0.1cm}\textrm{Pr}\left[\left.C^{\textrm{NL}}\right|X^{\textrm{L}}=x\right]\hspace{-0.1cm}f_{X}^{\textrm{L}}\left(x\right)dx\nonumber \\
\hspace{-0.3cm} & = & \hspace{-0.3cm}\int_{r}^{+\infty}\textrm{Pr}\left[\left.C^{\textrm{NL}}\right|X^{\textrm{L}}=x\right]f_{X}^{\textrm{L}}\left(x\right)dx,\label{eq:CCDF_SINR_req_UAS1_LoS}
\end{eqnarray}
where $\mathbb{E}_{\left[X\right]}\left\{ \cdot\right\} $
in the step (a) of (\ref{eq:CCDF_SINR_req_UAS1_LoS}) denotes the
expectation operation taking the expectation over the RV $X$
and the step (b) of (\ref{eq:CCDF_SINR_req_UAS1_LoS}) is valid because
(i) when $0<x\leq r$, it is apparent that $\textrm{Pr}\left[\left.R^{\textrm{L}}>r\right|X^{\textrm{L}}=x\right]=0$;
and (ii) when $x>r$, the conditional event $\left[\left.R^{\textrm{L}}>r\right|X^{\textrm{L}}=x\right]$
is equivalent to the conditional event $\left[\left.C^{\textrm{NL}}\right|X^{\textrm{L}}=x\right]$
. Taking the derivative of $\left(1-\bar{F}_{R}^{\textrm{L}}\left(r\right)\right)$
with regard to $r$, we can get the PDF of $R^{\textrm{L}}$ as
\begin{equation}
f_{R}^{\textrm{L}}\left(r\right)=\textrm{Pr}\left[\left.C^{\textrm{NL}}\right|X^{\textrm{L}}=r\right]f_{X}^{\textrm{L}}\left(r\right).\label{eq:geom_dis_PDF_total_UAS1_LoS}
\end{equation}
Considering the distance range of $d_{n-1}<r\leq d_{n}$,
we can extract the segment of $f_{R,n}^{\textrm{L}}\left(r\right)$
from $f_{R}^{\textrm{L}}\left(r\right)$ as
\begin{eqnarray}
f_{R,n}^{\textrm{L}}\left(r\right)\hspace{-0.3cm} & = & \hspace{-0.3cm}\exp\left(-\int_{0}^{r_{1}}\left(1-\textrm{Pr}^{\textrm{L}}\left(u\right)\right)2\pi u\lambda du\right)\nonumber \\
\hspace{-0.3cm} &  & \hspace{-0.3cm}\times\exp\left(-\int_{0}^{r}\textrm{Pr}^{\textrm{L}}\left(u\right)2\pi u\lambda du\right)\nonumber \\
\hspace{-0.3cm} &  & \hspace{-0.3cm}\times\textrm{Pr}_{n}^{\textrm{L}}\left(r\right)2\pi r\lambda,\quad\left(d_{n-1}<r\leq d_{n}\right),\label{eq:geom_dis_PDF_UAS1_LoS}
\end{eqnarray}
where $r_{1}=\underset{r_{1}}{\arg}\left\{ \zeta^{\textrm{NL}}\left(r_{1}\right)=\zeta_{n}^{\textrm{L}}\left(r\right)\right\} $.

Having obtained $f_{R,n}^{\textrm{L}}\left(r\right)$, we move on
to evaluate $\textrm{Pr}\left[\frac{P\zeta_{n}^{\textrm{L}}\left(r\right)h}{I_{r}+N_{0}}>\gamma\right]$
in~(\ref{eq:p_cov_general_form}) as
\begin{eqnarray}
\textrm{Pr}\left[\frac{P\zeta_{n}^{\textrm{L}}\left(r\right)h}{I_{r}+N_{0}}>\gamma\right]\hspace{-0.3cm} & = & \hspace{-0.3cm}\mathbb{E}_{\left[I_{r}\right]}\left\{ \textrm{Pr}\left[h>\frac{\gamma\left(I_{r}+N_{0}\right)}{P\zeta_{n}^{\textrm{L}}\left(r\right)}\right]\right\} \nonumber \\
\hspace{-0.3cm} & = & \hspace{-0.3cm}\mathbb{E}_{\left[I_{r}\right]}\left\{ \bar{F}_{H}\left(\frac{\gamma\left(I_{r}+N_{0}\right)}{P\zeta_{n}^{\textrm{L}}\left(r\right)}\right)\right\} ,\label{eq:Pr_SINR_req_UAS1_LoS}
\end{eqnarray}
where $\bar{F}_{H}\left(h\right)$ denotes the CCDF of RV
$h$. Since we assume $h$ to be an exponential RV, we have $\bar{F}_{H}\left(h\right)=\exp\left(-h\right)$
and thus~(\ref{eq:Pr_SINR_req_UAS1_LoS}) can be further derived as
\begin{eqnarray}
\hspace{-0.3cm} & \hspace{-0.3cm} & \hspace{-0.3cm}\hspace{-0.3cm}\hspace{-0.3cm}\hspace{-0.3cm}\textrm{Pr}\left[\frac{P\zeta_{n}^{\textrm{L}}\left(r\right)h}{I_{r}+N_{0}}>\gamma\right]\nonumber \\
\hspace{-0.3cm} & = & \hspace{-0.3cm}\mathbb{E}_{\left[I_{r}\right]}\left\{ \exp\left(-\frac{\gamma\left(I_{r}+N_{0}\right)}{P\zeta_{n}^{\textrm{L}}\left(r\right)}\right)\right\} \nonumber \\
\hspace{-0.3cm} & = & \hspace{-0.3cm}\exp\left(-\frac{\gamma N_{0}}{P\zeta_{n}^{\textrm{L}}\left(r\right)}\right)\mathbb{E}_{\left[I_{r}\right]}\left\{ \exp\left(-\frac{\gamma}{P\zeta_{n}^{\textrm{L}}\left(r\right)}I_{r}\right)\right\} \nonumber \\
\hspace{-0.3cm} & = & \hspace{-0.3cm}\exp\left(-\frac{\gamma N_{0}}{P\zeta_{n}^{\textrm{L}}\left(r\right)}\right)\mathscr{L}_{I_{r}}\left(\frac{\gamma}{P\zeta_{n}^{\textrm{L}}\left(r\right)}\right),\label{eq:Pr_SINR_req_wLT_UAS1_LoS}
\end{eqnarray}

\noindent where $\mathscr{L}_{I_{r}}\left(s\right)$ is the Laplace
transform of $I_{r}$ evaluated at $s$.

Next, we discuss the computation of $f_{R,n}^{\textrm{NL}}\left(r\right)$
in~(\ref{eq:p_cov_general_form}). Similar to the process to obtain
$f_{R,n}^{\textrm{L}}\left(r\right)$, we also define two events as
follows.%
\begin{comment}
the joint event of which is equivalent to the event that the UE is
associated with a BS with a LoS path at distance $r$ according to
UAS~1.
\end{comment}

\begin{itemize}
\item \textbf{Event $B^{\textrm{NL}}$: The nearest BS with a NLoS path to the typical UE,
is located at distance $X^{\textrm{NL}}$.} Similar to (\ref{eq:PDF_X_BL}),
the PDF of $X^{\textrm{NL}}$ is given by
\begin{eqnarray}
f_{X}^{\textrm{NL}}\left(x\right)\hspace{-0.1cm} & = & \hspace{-0.1cm}\exp\left(-\int_{0}^{x}\left(1-\textrm{Pr}^{\textrm{L}}\left(u\right)\right)2\pi u\lambda du\right)\nonumber \\
\hspace{-0.1cm} &  & \hspace{-0.1cm}\times\left(1-\textrm{Pr}^{\textrm{L}}\left(x\right)\right)2\pi x\lambda.\label{eq:PDF_X_BNL}
\end{eqnarray}

\item \textbf{Event $C^{\textrm{L}}$ conditioned on the value of $X^{\textrm{NL}}$:
Given that $X^{\textrm{NL}}=x$, the typical UE is then associated with such BS at distance $X^{\textrm{NL}}=x$.}
To make the typical UE associated with the NLoS BS at distance $X^{\textrm{NL}}=x$, such BS should give the smallest path loss (i.e., the largest $\zeta\left(r\right)$) from such BS to the typical UE, i.e.,
there should be no BS with a LoS path inside the disk centered on the UE with a radius of $x_{2}<x$,
where $x_{2}$ satisfies $x_{2}=\underset{x_{2}}{\arg}\left\{ \zeta^{\textrm{L}}\left(x_{2}\right)=\zeta^{\textrm{NL}}\left(x\right)\right\}$.
Similar to (\ref{eq:cond_prob_CNL}), such conditional probability
of $C^{\textrm{L}}$ on condition of $X^{\textrm{NL}}=x$ can be written as
\begin{equation}
\hspace{-0.1cm}\hspace{-0.1cm}\textrm{Pr}\left[\left.C^{\textrm{L}}\right|X^{\textrm{NL}}=x\right]=\exp\left(\hspace{-0.1cm}-\hspace{-0.1cm}\int_{0}^{x_{2}}\hspace{-0.1cm}\hspace{-0.1cm}\textrm{Pr}^{\textrm{L}}\left(u\right)2\pi u\lambda du\right)\hspace{-0.1cm}.\label{eq:cond_prob_CL}
\end{equation}

\end{itemize}

Then, we consider an unconditional event that the typical UE is associated with
a BS with a NLoS path and such BS is located at distance $R^{\textrm{NL}}$.
Similar to (\ref{eq:CCDF_SINR_req_UAS1_LoS}), the CCDF of $R^{\textrm{NL}}$,
denoted by $\bar{F}_{R}^{\textrm{NL}}\left(r\right)$, can be derived
as
\begin{eqnarray}
\bar{F}_{R}^{\textrm{NL}}\left(r\right)\hspace{-0.1cm} & = & \hspace{-0.1cm}\textrm{Pr}\left[R^{\textrm{NL}}>r\right]\nonumber \\
\hspace{-0.1cm} & = & \hspace{-0.1cm}\int_{r}^{+\infty}\textrm{Pr}\left[\left.C^{\textrm{L}}\right|X^{\textrm{NL}}=x\right]f_{X}^{\textrm{NL}}\left(x\right)dx.\label{eq:CCDF_SINR_req_UAS1_NLoS}
\end{eqnarray}
Taking the derivative of $\left(1-\bar{F}_{R}^{\textrm{NL}}\left(r\right)\right)$
with regard to $r$, we can get the PDF of $R^{\textrm{NL}}$ as
\begin{equation}
f_{R}^{\textrm{NL}}\left(r\right)=\textrm{Pr}\left[\left.C^{\textrm{L}}\right|X^{\textrm{NL}}=r\right]f_{X}^{\textrm{NL}}\left(x\right).\label{eq:geom_dis_PDF_total_UAS1_NLoS}
\end{equation}

\noindent Considering the distance range of $d_{n-1}<r\leq d_{n}$,
we can extract the segment of $f_{R,n}^{\textrm{NL}}\left(r\right)$
from $f_{R}^{\textrm{NL}}\left(r\right)$ as
\begin{eqnarray}
\hspace{-0.3cm}\hspace{-0.3cm}f_{R,n}^{\textrm{NL}}\left(r\right)\hspace{-0.3cm} & = & \hspace{-0.3cm}\exp\left(-\int_{0}^{r_{2}}\textrm{Pr}^{\textrm{L}}\left(u\right)2\pi u\lambda du\right)\nonumber \\
\hspace{-0.3cm} &  & \hspace{-0.3cm}\times\exp\left(-\int_{0}^{r}\left(1-\textrm{Pr}^{\textrm{L}}\left(u\right)\right)2\pi u\lambda du\right)\nonumber \\
\hspace{-0.3cm} &  & \hspace{-0.3cm}\times\left(1-\textrm{Pr}_{n}^{\textrm{L}}\left(r\right)\right)2\pi r\lambda,\quad\left(d_{n-1}<r\leq d_{n}\right),\label{eq:geom_dis_PDF_UAS1_NLoS}
\end{eqnarray}

\noindent where $r_{2}=\underset{r_{2}}{\arg}\left\{ \zeta^{\textrm{L}}\left(r_{2}\right)=\zeta_{n}^{\textrm{NL}}\left(r\right)\right\} $.

Similar to (\ref{eq:Pr_SINR_req_wLT_UAS1_LoS}), $\textrm{Pr}\left[\frac{P\zeta_{n}^{\textrm{NL}}\left(r\right)h}{I_{r}+N_{0}}>\gamma\right]$
can be computed by
\begin{eqnarray}
\hspace{-0.1cm}\hspace{-0.1cm}\hspace{-0.1cm}\hspace{-0.1cm}\hspace{-0.1cm}\hspace{-0.1cm}\textrm{Pr}\left[\frac{P\zeta_{n}^{\textrm{NL}}\left(r\right)h}{I_{r}+N_{0}}>\gamma\right]\hspace{-0.3cm} & = & \hspace{-0.3cm}\mathbb{E}_{\left[I_{r}\right]}\left\{ \exp\left(-\frac{\gamma\left(I_{r}+N_{0}\right)}{P\zeta_{n}^{\textrm{NL}}\left(r\right)}\right)\right\} \nonumber \\
\hspace{-0.3cm} & = & \hspace{-0.3cm}\exp\hspace{-0.1cm}\left(\hspace{-0.1cm}-\frac{\gamma N_{0}}{P\zeta_{n}^{\textrm{NL}}\left(r\right)}\hspace{-0.1cm}\right)\hspace{-0.1cm}\mathscr{L}_{I_{r}}\hspace{-0.1cm}\left(\hspace{-0.1cm}\frac{\gamma}{P\zeta_{n}^{\textrm{NL}}\left(r\right)}\hspace{-0.1cm}\right)\hspace{-0.1cm}.\label{eq:Pr_SINR_req_wLT_UAS1_NLoS}
\end{eqnarray}

Our proof of Theorem~\ref{thm:p_cov_UAS1} is completed by plugging
(\ref{eq:geom_dis_PDF_UAS1_LoS}), (\ref{eq:Pr_SINR_req_wLT_UAS1_LoS}),
(\ref{eq:geom_dis_PDF_UAS1_NLoS}) and (\ref{eq:Pr_SINR_req_wLT_UAS1_NLoS})
into (\ref{eq:p_cov_general_form}).

\section*{Appendix~B: Proof of Lemma~\ref{lem:laplace_term_UAS1_LoS_seg1}\label{sec:Appendix-B}}

Based on the considered UAS, it is straightforward to derive $\mathscr{L}_{I_{r}}\left(s\right)$
in the range of $0<r\leq d_{1}$ as
\begin{eqnarray}
\hspace{-0.3cm} & \hspace{-0.3cm} & \hspace{-0.3cm}\hspace{-0.3cm}\hspace{-0.1cm}\hspace{-0.1cm}\mathscr{L}_{I_{r}}\left(s\right)\nonumber \\
\hspace{-0.3cm} & = & \hspace{-0.3cm}\mathbb{E}_{\left[I_{r}\right]}\left\{ \left.\exp\left(-sI_{r}\right)\right|0<r\leq d_{1}\right\} \nonumber \\
\hspace{-0.3cm} & = & \hspace{-0.3cm}\mathbb{E}_{\left[\Phi,\left\{ \beta_{i}\right\} ,\left\{ g_{i}\right\} \right]}\left\{ \left.\exp\left(-s\sum_{i\in\Phi/b_{o}}P\beta_{i}g_{i}\right)\right|0<r\leq d_{1}\right\} \nonumber \\
\hspace{-0.3cm} & \overset{(a)}{=} & \hspace{-0.3cm}\exp\hspace{-0.1cm}\left(\hspace{-0.1cm}\left.-2\pi\lambda\hspace{-0.1cm}\int_{r}^{\infty}\hspace{-0.1cm}\hspace{-0.1cm}\hspace{-0.1cm}\left(1\hspace{-0.1cm}-\hspace{-0.1cm}\mathbb{E}_{\left[g\right]}\hspace{-0.1cm}\left\{ \exp\left(-sP\beta\left(u\right)g\right)\right\} \right)\hspace{-0.1cm}udu\right|\hspace{-0.1cm}0<r\leq d_{1}\hspace{-0.1cm}\right)\hspace{-0.1cm},\nonumber \\
\hspace{-0.3cm} & \hspace{-0.3cm}\label{eq:laplace_term_LoS_UAS1_seg1_proof_eq1}
\end{eqnarray}
where the step (a) of (\ref{eq:laplace_term_LoS_UAS1_seg1_proof_eq1})
is obtained from~\cite{Jeff's work 2011}.

Since $0<r\leq d_{1}$, $\mathbb{E}_{\left[g\right]}\left\{ \exp\left(-sP\beta\left(u\right)g\right)\right\} $
in (\ref{eq:laplace_term_LoS_UAS1_seg1_proof_eq1}) should consider
interference from both LoS and NLoS paths. Thus, $\mathscr{L}_{I_{r}}\left(s\right)$
can be further derived as
\begin{eqnarray}
\hspace{-0.3cm} & \hspace{-0.3cm} & \hspace{-0.3cm}\hspace{-0.3cm}\hspace{-0.1cm}\hspace{-0.1cm}\mathscr{L}_{I_{r}}\left(s\right)\nonumber \\
 & = & \hspace{-0.3cm}\exp\left(-2\pi\lambda\int_{r}^{d_{1}}\left(1-\frac{u}{d_{1}}\right)\right.\nonumber \\
 &  & \hspace{-0.3cm}\hspace{-0.3cm}\left.{\color{white}\int_{dummy}^{dummy}}\times\left[1-\mathbb{E}_{\left[g\right]}\left\{ \exp\left(-sPA^{{\rm {L}}}u^{-\alpha^{{\rm {L}}}}g\right)\right\} \right]udu\right)\nonumber \\
\hspace{-0.3cm} &  & \hspace{-0.3cm}\times\exp\left(-2\pi\lambda\int_{r_{1}}^{d_{1}}\frac{u}{d_{1}}\right.\nonumber \\
 &  & \hspace{-0.3cm}\hspace{-0.3cm}\left.{\color{white}\int_{dummy}^{dummy}}\times\left[1-\mathbb{E}_{\left[g\right]}\left\{ \exp\left(-sPA^{{\rm {NL}}}u^{-\alpha^{{\rm {NL}}}}g\right)\right\} \right]udu\right)\nonumber \\
\hspace{-0.3cm} &  & \hspace{-0.3cm}\times\exp\left(-2\pi\lambda\int_{d_{1}}^{\infty}1\right.\nonumber \\
 &  & \hspace{-0.3cm}\hspace{-0.3cm}\left.{\color{white}\int_{dummy}^{dummy}}\times\left[1-\mathbb{E}_{\left[g\right]}\left\{ \exp\left(-sPA^{{\rm {NL}}}u^{-\alpha^{{\rm {NL}}}}g\right)\right\} \right]udu\right)\nonumber \\
\hspace{-0.3cm} & = & \hspace{-0.3cm}\exp\left(-2\pi\lambda\int_{r}^{d_{1}}\left(1-\frac{u}{d_{1}}\right)\frac{u}{1+\left(sPA^{{\rm {L}}}\right)^{-1}u^{\alpha^{{\rm {L}}}}}du\right)\nonumber \\
\hspace{-0.3cm} &  & \hspace{-0.3cm}\times\exp\left(-2\pi\lambda\int_{r_{1}}^{d_{1}}\frac{u}{d_{1}}\frac{u}{1+\left(sPA^{{\rm {NL}}}\right)^{-1}u^{\alpha^{{\rm {NL}}}}}du\right)\nonumber \\
\hspace{-0.3cm} &  & \hspace{-0.3cm}\times\exp\left(-2\pi\lambda\int_{d_{1}}^{\infty}\frac{u}{1+\left(sPA^{{\rm {NL}}}\right)^{-1}u^{\alpha^{{\rm {NL}}}}}du\right).\hspace{-0.3cm}\label{eq:laplace_term_LoS_UAS1_seg1_proof_eq2}
\end{eqnarray}

Plugging $s=\frac{\gamma r^{\alpha^{\textrm{L}}}}{PA^{\textrm{L}}}$
into (\ref{eq:laplace_term_LoS_UAS1_seg1_proof_eq2}), and with some mathematical manipulations considering the definition of $\rho_{1}\left(\alpha,\beta,t,d\right)$ and $\rho_{2}\left(\alpha,\beta,t,d\right)$
in (\ref{eq:rou1_func}) and (\ref{eq:rou2_func}), we can obtain
$\mathscr{L}_{I_{r}}\left(\frac{\gamma r^{\alpha^{{\rm {L}}}}}{PA^{{\rm {L}}}}\right)$
shown in (\ref{eq:Lemma_3}), which concludes our proof.

\section*{\noindent Appendix~C: Proof of Lemma~\ref{lem:laplace_term_UAS1_NLoS_seg1}\label{sec:Appendix-C}}

Following the recipe employed in Appendix~B, we consider interference from both LoS and NLoS paths, and we can derive $\mathscr{L}_{I_{r}}\left(\frac{\gamma r^{\alpha^{{\rm {NL}}}}}{PA^{{\rm {NL}}}}\right)$
in the range of $0<r\leq y_{1}$ as
%\begin{eqnarray}
%\hspace{-0.3cm} & \hspace{-0.3cm} & \hspace{-0.3cm}\hspace{-0.3cm}\hspace{-0.1cm}\hspace{-0.1cm}\mathscr{L}_{I_{r}}\left(\frac{\gamma r^{\alpha^{{\rm {NL}}}}}{PA^{{\rm {NL}}}}\right)\nonumber \\
%\hspace{-0.3cm} & = & \hspace{-0.3cm}\exp\hspace{-0.1cm}\left(-2\pi\lambda\int_{r_{2}}^{d_{1}}\hspace{-0.2cm}\left(1-\frac{u}{d_{1}}\right)\frac{u}{1+\left(\frac{\gamma r^{\alpha^{{\rm {NL}}}}}{PA^{{\rm {NL}}}}PA^{{\rm {L}}}\right)^{-1}u^{\alpha^{{\rm {L}}}}}du\right)\nonumber \\
%\hspace{-0.3cm} &  & \hspace{-0.3cm}\times\exp\hspace{-0.1cm}\left(-2\pi\lambda\int_{r}^{d_{1}}\hspace{-0.2cm}\frac{u}{d_{1}}\frac{u}{1+\left(\frac{\gamma r^{\alpha^{{\rm {NL}}}}}{PA^{{\rm {NL}}}}PA^{{\rm {NL}}}\right)^{-1}u^{\alpha^{{\rm {NL}}}}}du\right)\nonumber \\
%\hspace{-0.3cm} &  & \hspace{-0.3cm}\times\exp\hspace{-0.1cm}\left(-2\pi\lambda\int_{d_{1}}^{\infty}\hspace{-0.2cm}\frac{u}{1+\left(\frac{\gamma r^{\alpha^{{\rm {NL}}}}}{PA^{{\rm {NL}}}}PA^{{\rm {NL}}}\right)^{-1}u^{\alpha^{{\rm {NL}}}}}du\right),\nonumber \\
%\hspace{-0.3cm} &  & \hspace{-0.3cm}\hspace{5.5cm}\left(0<r\leq y_{1}\right).\label{eq:proof_Lemma4_eq1}
%\end{eqnarray}
\begin{eqnarray}
\hspace{-0.3cm} & \hspace{-0.3cm} & \hspace{-0.3cm}\hspace{-0.3cm}\hspace{-0.3cm}\mathscr{L}_{I_{r}}\left(\frac{\gamma r^{\alpha^{{\rm {NL}}}}}{PA^{{\rm {NL}}}}\right)\nonumber \\
 & = & \hspace{-0.3cm}\exp\hspace{-0.1cm}\left(\hspace{-0.1cm}-2\pi\lambda\int_{r_{2}}^{d_{1}}\left(1-\frac{u}{d_{1}}\right)\right.\nonumber \\
 &  & \hspace{-0.3cm}\hspace{-0.3cm}\hspace{-0.3cm}\hspace{-0.3cm}\left.{\color{white}\int_{dummy}^{dummy}}\times\left[1-\mathbb{E}_{\left[g\right]}\left\{ \hspace{-0.1cm}\exp\left(-\frac{\gamma r^{\alpha^{{\rm {NL}}}}}{PA^{{\rm {NL}}}}PA^{{\rm {L}}}u^{-\alpha^{{\rm {L}}}}g\right)\hspace{-0.1cm}\right\} \right]\hspace{-0.1cm}udu\hspace{-0.1cm}\right)\nonumber \\
\hspace{-0.3cm} &  & \hspace{-0.3cm}\times\exp\hspace{-0.1cm}\left(\hspace{-0.1cm}-2\pi\lambda\int_{r}^{d_{1}}\frac{u}{d_{1}}\right.\nonumber \\
 &  & \hspace{-0.3cm}\hspace{-0.3cm}\hspace{-0.3cm}\hspace{-0.3cm}\left.{\color{white}\int_{dummy}^{dummy}}\times\left[1-\mathbb{E}_{\left[g\right]}\left\{ \hspace{-0.1cm}\exp\left(-\frac{\gamma r^{\alpha^{{\rm {NL}}}}}{PA^{{\rm {NL}}}}PA^{{\rm {NL}}}u^{-\alpha^{{\rm {NL}}}}g\right)\hspace{-0.1cm}\right\} \right]\hspace{-0.1cm}udu\hspace{-0.1cm}\right)\nonumber \\
\hspace{-0.3cm} &  & \hspace{-0.3cm}\times\exp\hspace{-0.1cm}\left(\hspace{-0.1cm}-2\pi\lambda\int_{d_{1}}^{\infty}1\right.\nonumber \\
 &  & \hspace{-0.3cm}\hspace{-0.3cm}\hspace{-0.3cm}\hspace{-0.3cm}\left.{\color{white}\int_{dummy}^{dummy}}\times\left[1-\mathbb{E}_{\left[g\right]}\left\{ \hspace{-0.1cm}\exp\left(-\frac{\gamma r^{\alpha^{{\rm {NL}}}}}{PA^{{\rm {NL}}}}PA^{{\rm {NL}}}u^{-\alpha^{{\rm {NL}}}}g\right)\hspace{-0.1cm}\right\} \right]\hspace{-0.1cm}udu\hspace{-0.1cm}\right)\nonumber \\
\hspace{-0.3cm} & = & \hspace{-0.3cm}\exp\hspace{-0.1cm}\left(\hspace{-0.1cm}-2\pi\lambda\int_{r_{2}}^{d_{1}}\hspace{-0.2cm}\left(1-\frac{u}{d_{1}}\right)\hspace{-0.1cm}\frac{u}{1+\left(\frac{\gamma r^{\alpha^{{\rm {NL}}}}}{PA^{{\rm {NL}}}}PA^{{\rm {L}}}\right)^{-1}u^{\alpha^{{\rm {L}}}}}du\hspace{-0.1cm}\right)\nonumber \\
\hspace{-0.3cm} &  & \hspace{-0.3cm}\times\exp\hspace{-0.1cm}\left(\hspace{-0.1cm}-2\pi\lambda\int_{r}^{d_{1}}\hspace{-0.2cm}\frac{u}{d_{1}}\frac{u}{1+\left(\frac{\gamma r^{\alpha^{{\rm {NL}}}}}{PA^{{\rm {NL}}}}PA^{{\rm {NL}}}\right)^{-1}u^{\alpha^{{\rm {NL}}}}}du\hspace{-0.1cm}\right)\nonumber \\
\hspace{-0.3cm} &  & \hspace{-0.3cm}\times\exp\hspace{-0.1cm}\left(\hspace{-0.1cm}-2\pi\lambda\int_{d_{1}}^{\infty}\hspace{-0.2cm}\frac{u}{1+\left(\frac{\gamma r^{\alpha^{{\rm {NL}}}}}{PA^{{\rm {NL}}}}PA^{{\rm {NL}}}\right)^{-1}u^{\alpha^{{\rm {NL}}}}}du\hspace{-0.1cm}\right),\nonumber \\
\hspace{-0.3cm} &  & \hspace{-0.3cm}\hspace{5cm}\left(0<r\leq y_{1}\right).\label{eq:proof_Lemma4_eq1}
\end{eqnarray}

For $\mathscr{L}_{I_{r}}\left(\frac{\gamma r^{\alpha^{{\rm {NL}}}}}{PA^{{\rm {NL}}}}\right)$
in the range of $y_{1}<r\leq d_{1}$, we consider interference only from NLoS paths and we can derive it as
\begin{eqnarray}
\hspace{-0.3cm} & \hspace{-0.3cm} & \hspace{-0.3cm}\hspace{-0.3cm}\hspace{-0.1cm}\hspace{-0.1cm}\hspace{-0.1cm}\mathscr{L}_{I_{r}}\left(\frac{\gamma r^{\alpha^{{\rm {NL}}}}}{PA^{{\rm {NL}}}}\right)\nonumber \\
\hspace{-0.3cm} & = & \hspace{-0.3cm}\exp\left(-2\pi\lambda\int_{r}^{d_{1}}\frac{u}{d_{1}}\frac{u}{1+\left(\frac{\gamma r^{\alpha^{{\rm {NL}}}}}{PA^{{\rm {NL}}}}PA^{{\rm {NL}}}\right)^{-1}u^{\alpha^{{\rm {NL}}}}}du\right)\nonumber \\
\hspace{-0.3cm} &  & \hspace{-0.3cm}\times\exp\left(-2\pi\lambda\int_{d_{1}}^{\infty}\frac{u}{1+\left(\frac{\gamma r^{\alpha^{{\rm {NL}}}}}{PA^{{\rm {NL}}}}PA^{{\rm {NL}}}\right)^{-1}u^{\alpha^{{\rm {NL}}}}}du\right),\nonumber \\
\hspace{-0.3cm} &  & \hspace{-0.3cm}\hspace{5cm}\left(y_{1}<r\leq d_{1}\right).\label{eq:proof_Lemma4_eq2}
\end{eqnarray}

Our proof is thus completed by plugging (\ref{eq:rou1_func})
and (\ref{eq:rou2_func}) into (\ref{eq:proof_Lemma4_eq1}) and (\ref{eq:proof_Lemma4_eq2}).

\section*{Appendix~D: Proof of Lemma~\ref{lem:laplace_term_UAS1_NLoS_seg2}\label{sec:Appendix-D}}

Following the recipe employed in Appendix~B, we consider
interference only from NLoS paths and we can derive $\mathscr{L}_{I_{r}}\left(\frac{\gamma r^{\alpha^{{\rm {NL}}}}}{PA^{{\rm {NL}}}}\right)$
in the range of $r>d_{1}$ as
%\begin{eqnarray}
%\mathscr{L}_{I_{r}}\left(\frac{\gamma r^{\alpha^{{\rm {NL}}}}}{PA^{{\rm {NL}}}}\right)\hspace{-0.3cm} & = & \hspace{-0.3cm}\exp\left(-2\pi\lambda\int_{r}^{\infty}\frac{u}{1+\left(\gamma r^{\alpha^{{\rm {NL}}}}\right)^{-1}u^{\alpha^{{\rm {NL}}}}}du\right),\nonumber \\
%\hspace{-0.3cm} & \hspace{-0.3cm} & \hspace{4cm}\left(r>d_{1}\right).\label{eq:proof_Lemma5_eq1}
%\end{eqnarray}
\begin{eqnarray}
\hspace{-0.3cm} & \hspace{-0.3cm} & \hspace{-0.3cm}\hspace{-0.3cm}\hspace{-0.3cm}\mathscr{L}_{I_{r}}\left(\frac{\gamma r^{\alpha^{{\rm {NL}}}}}{PA^{{\rm {NL}}}}\right)\nonumber \\
\hspace{-0.3cm} & = & \hspace{-0.3cm}\exp\hspace{-0.1cm}\left(\hspace{-0.1cm}-2\pi\lambda\int_{r}^{\infty}1\right.\nonumber \\
 &  & \hspace{-0.3cm}\hspace{-0.3cm}\hspace{-0.3cm}\hspace{-0.3cm}\left.{\color{white}\int_{dummy}^{dummy}}\times\left[1-\mathbb{E}_{\left[g\right]}\left\{ \hspace{-0.1cm}\exp\left(-\frac{\gamma r^{\alpha^{{\rm {NL}}}}}{PA^{{\rm {NL}}}}PA^{{\rm {NL}}}u^{-\alpha^{{\rm {NL}}}}g\right)\hspace{-0.1cm}\right\} \right]\hspace{-0.1cm}udu\hspace{-0.1cm}\right)\nonumber \\
\hspace{-0.3cm} & = & \hspace{-0.3cm}\exp\hspace{-0.1cm}\left(\hspace{-0.1cm}-2\pi\lambda\int_{r}^{\infty}\hspace{-0.2cm}\frac{u}{1+\left(\gamma r^{\alpha^{{\rm {NL}}}}\right)^{-1}u^{\alpha^{{\rm {NL}}}}}du\hspace{-0.1cm}\right),\left(r>d_{1}\right).\label{eq:proof_Lemma5_eq1}
\end{eqnarray}

Our proof is thus completed by plugging (\ref{eq:rou2_func})
into (\ref{eq:proof_Lemma5_eq1}).

\begin{IEEEbiography}[{\includegraphics[clip,width=1in]{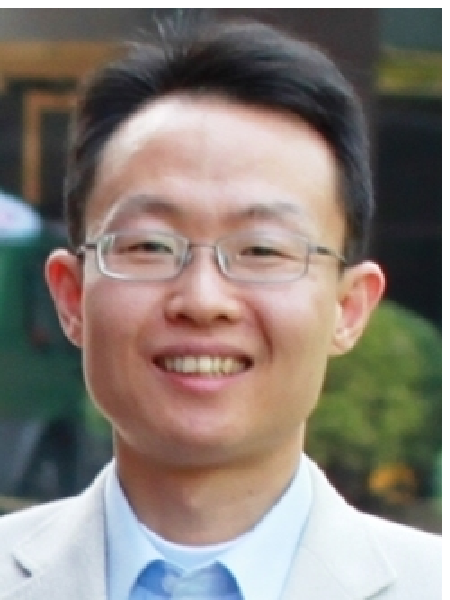}}]
{Ming Ding} (M'12) is a researcher at Data 61, Australia. He received his B.S. and M.S. degrees with first class honours in Electronics Engineering from Shanghai Jiao Tong University (SJTU), China, in 2004 and 2007, respectively. In Apr. 2007, he joined Sharp Laboratories of China (SLC) as a Researcher. From Sep. 2007 to Sep. 2011, he pursued his Doctor in Philosophy (Ph.D.) at SJTU while at the same time working as a Researcher/Senior Researcher at SLC. In Dec. 2011, he achieved his Ph.D. in Signal and Information Processing from SJTU and continued to work for SLC as a Senior Researcher/Principal Researcher until Sep. 2014 when he joined National Information and Communications Technology Australia (NICTA). In Sep. 2015, Commonwealth Scientific and Industrial Research Organization (CSIRO) and NICTA joined forces to create Data 61, where he continued his research in this new R{\&}D center located in Australia. Ming has been working on B3G, 4G and 5G wireless communication networks for more than 9 years and his research interests include synchronization, MIMO technology, cooperative communications, heterogeneous networks, device-to-device communications, and modelling of wireless communication systems. Besides, he served as the Algorithm Design Director and Programming Director for a system-level simulator of future telecommunication networks in SLC for more than 7 years. Up to now, Ming has published more than 30 papers in IEEE journals and conferences, all in recognized venues, and about 20 3GPP standardization contributions, as well as a Springer book "Multi-point Cooperative Communication Systems: Theory and Applications". Also, as the first inventor, he holds 8 CN, 2 JP, 2 KR patents and filed another 30 patent applications on 4G/5G technologies. For his inventions and publications, he was the recipient of the President's Award of SLC in 2012 and served as one of the key members in the 4G/5G standardization team when it was awarded in 2014 as Sharp Company Best Team: LTE Standardization Patent Portfolio. Ming is or has been guest editor/co-chair/TPC member of several IEEE top-tier journals/conferences, e.g., IEEE JSAC, IEEE Comm. Mag., IEEE Globecom workshops.
\end{IEEEbiography}

\vspace{-1cm}

\begin{IEEEbiography}[{\includegraphics[clip,width=1in]{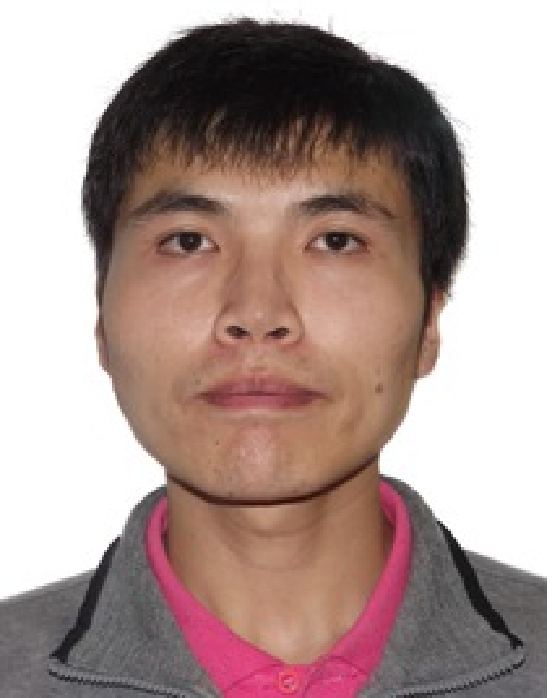}}]
{Peng Wang} received the B.Sc. degree in applied electronics from Beijing University of Aeronautics and Astronautics, Beijing, China, in 2009, and the M.Eng. degree in telecommunications in 2012 from the Australian National University, Canberra, ACT, Australia. He is currently working towards the PhD degree in Engineering at The University of Sydney. He is also with Data 61, Australia. His research interests include wireless broadcast networks, heterogeneous networks, graph theory, and its application in networking, channel/network coding, 5G cellular systems, etc.
\end{IEEEbiography}

\vspace{-1cm}

\begin{IEEEbiography}[{\includegraphics[clip,width=1in]{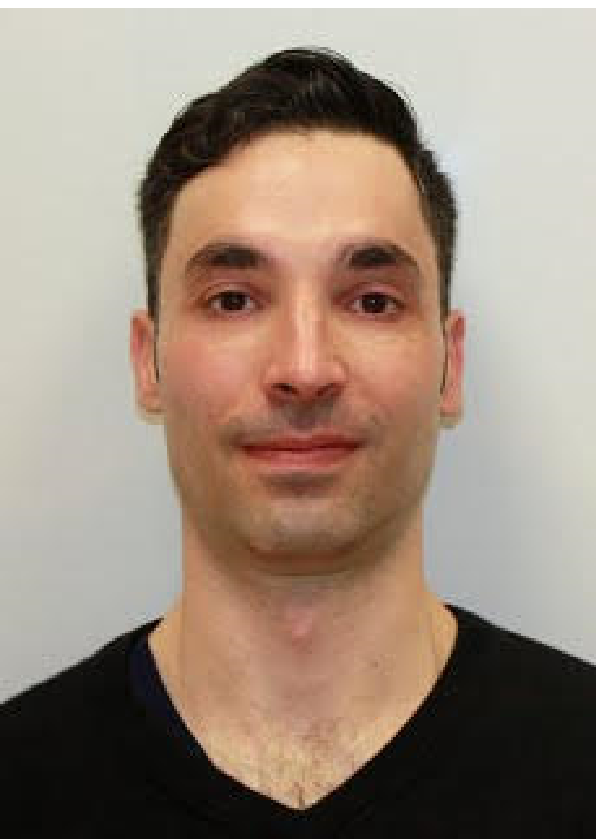}}]
{David L$\acute{\textrm{o}}$pez-P$\acute{\textrm{e}}$rez} is a Member of Technical Staff at Bell Laboratories, Alcatel-Lucent, and his main research interests are in HetNets, small cells, interference and mobility management as well as network optimization and simulation. Prior to this, David earned his PhD in Wireless Networking from the University of Bedfordshire, UK in Apr. '11. David was Research Associate at King's College London, UK from Aug. '10 to Dec. '11, carrying post-doctoral studies, and was with VODAFONE, Spain from Feb. '05 to Feb. '06, working in the area of network planning and optimization. David was also invited researcher at DOCOMO USA labs, CA in 2011, and CITI INSA, France in 2009. For his publications and patent contributions, David is a recipient of both the Bell Labs Alcatel-Lucent Award of Excellence and Certificate of Outstanding Achievement. He was also finalist for the Scientist of the Year prize in The Irish Laboratory Awards (2013). David has also been awarded as PhD Marie-Curie Fellow in 2007 and Exemplary Reviewer for IEEE Communications Letters in 2011. David is editor of the book "Heterogeneous Cellular Networks: Theory, Simulation and Deployment" Cambridge University Press, 2012. Moreover, he has published more than 70 book chapters, journal and conference papers, all in recognized venues, and filed more than 30 patents applications. David is or has been guest editor of a number of journals, e.g., IEEE JSAC, IEEE Comm. Mag., TPC member of top tier conferences, e.g., IEEE Globecom and IEEE PIMRC, and co-chair of a number of workshops.
\end{IEEEbiography}

\vspace{-1cm}

\begin{IEEEbiography}[{\includegraphics[clip,width=1in]{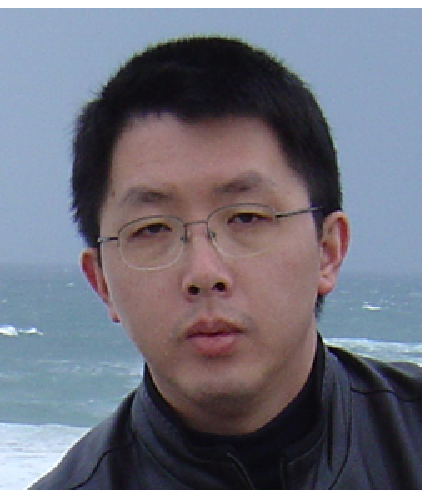}}]
{Guoqiang Mao} (M'02-SM'08) received PhD in telecommunications engineering in 2002 from Edith Cowan University. He currently holds the position of Professor of Wireless Networking, Director of Center for Real-time Information Networks at the University of Technology, Sydney. He has published more than 100 papers in international conferences and journals, which have been cited more than 3000 times. His research interest includes intelligent transport systems, applied graph theory and its applications in telecommunications, wireless sensor networks, wireless localization techniques and network performance analysis.
\end{IEEEbiography}

\vspace{-16cm}

\begin{IEEEbiography}[{\includegraphics[clip,width=1in]{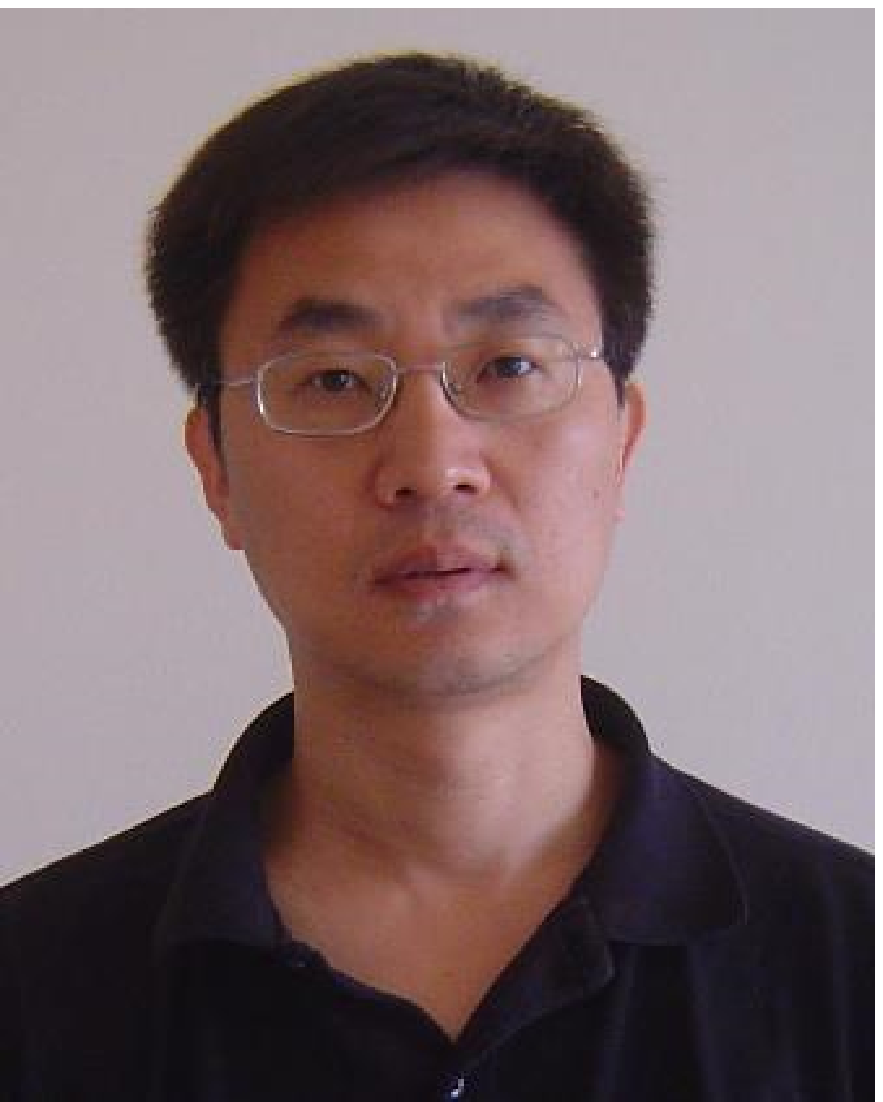}}]
{Zihuai Lin} received the Ph.D. degree in Electrical Engineering from Chalmers University of Technology, Sweden, in 2006. Prior to this he has held positions at Ericsson Research, Stockholm, Sweden. Following Ph.D. graduation, he worked as a Research Associate Professor at Aalborg University, Denmark and currently at the School of Electrical and Information Engineering, the University of Sydney, Australia. His research interests include source/channel/network coding, coded modulation, MIMO, OFDMA, SC-FDMA, radio resource management, cooperative communications, small-cell networks, 5G cellular systems, etc.
\end{IEEEbiography}

\end{document}